# High-resolution remote thermography using luminescent low-dimensional tin-halide perovskites

*Sergii Yakunin,[*,1,2] Bogdan M. Benin,[1,2] Yevhen Shynkarenko,[1,2] Olga Nazarenko,[1,2] Maryna I. Bodnarchuk,[1,2] Dmitry N. Dirin,[1, 2] Christoph Hofer,[3] Stefano Cattaneo[3] and Maksym V. Kovalenko[*1,2]*

[1] Laboratory of Inorganic Chemistry, Department of Chemistry and Applied Biosciences, ETH Zürich, CH-8093 Zürich, Switzerland

[2] Laboratory for Thin Films and Photovoltaics, Empa – Swiss Federal Laboratories for Materials Science and Technology, CH-8600 Dübendorf, Switzerland

[3] Swiss Center for Electronics and Microtechnology (CSEM), Center Landquart, CH-7302 Landquart, Switzerland

[*]E-mail: mvkovalenko@ethz.ch; yakunins@ethz.ch

While metal-halide perovskites have recently revolutionized research in optoelectronics through a unique combination of performance and synthetic simplicity, their low-dimensional counterparts can further expand the field with hitherto unknown and practically useful optical functionalities. In this context, we present the strong temperature dependence of the photoluminescence (PL) lifetime of low-dimensional, perovskite-like tin-halides, and apply this property to thermal imaging with a high precision of 0.05 °C. The PL lifetimes are governed by the heat-assisted de-trapping of self-trapped excitons, and their values can be varied over several orders of magnitude by adjusting the temperature (up to 20 ns °C$^{-1}$). Typically, this sensitive range spans up to one hundred centigrade, and it is both compound-specific and shown to be compositionally and structurally tunable from -100 to 110 ºC going from $[C(NH_2)_3]_2SnBr_4$ to $Cs_4SnBr_6$ and $(C_4N_2H_{14}I)_4SnI_6$. Finally, through the innovative implementation of cost-effective hardware for fluorescence lifetime imaging (FLI), based on time-of-flight (ToF) technology, these novel thermoluminophores have been used to record thermographic videos with high spatial and thermal resolution.

**Main.** Remote thermal imaging or thermography, lends itself to numerous applications ranging from medicine[1] and defense, to biological research[2], or the diagnosis of technical failures[3]. In all these applications, remote thermal detection falls into two main categories: infrared (IR) or visible. IR-based detectors exploit the long-wave IR emission (LWIR, *e.g.* thermal emission) of a studied object using the fact that the integral radiation intensity emitted from the blackbody scales with its temperature ($T$) as $\sim T^4$. This emission can be recorded with rather costly photodetector arrays composed of narrow bandgap semiconductors (InSb, $In_{1-x}Ga_xAs$ and $Hg_{1-x}Cd_xTe$)[4,5], and this method is predominantly used for scientific and military purposes. Alternative and less expensive bolometric detectors, owing to advancements in MEMS-technologies (Micro-Electro-Mechanical Systems) [6,7], have already entered the consumer electronics market, and are able to record thermal images with both high speed and high resolution. However, their thermographic performance, based on measurements of IR radiation intensity, is inherently limited by the transparency and emissivity/reflectivity of an observed object and, more importantly, by any material and medium (window, coating, matrix, solvent *etc.*) situated within the path between the detector and an object (Fig. 1). As one of the major consequences, IR thermography cannot be easily combined with conventional optical microscopy or other enclosed optical systems such as cryostats or microfluidic cells.

An alternative method for remote thermography, which is unhindered by enclosures or IR-absorptive media, utilizes temperature sensitive luminophores (i.e. fluorophores or phosphors) with PL in the visible spectral range (Fig. 1c) that are deposited onto, or incorporated into, the object of interest as temperature probes[8-16]. To probe an object's temperature, the luminophore is then excited by an ultraviolet or visible (UV-Vis) pulsed source (*e.g.* laser or light-emitting diode) and the temperature-dependent PL lifetime decay is then analyzed by time-resolving detectors. This PL-lifetime approach exhibits several benefits: the excitation power and, consequently the PL intensity, can be adjusted to a value appropriate for the dynamic range of the detector. Additionally, the use of UV-Vis light, rather than mid- to long-wavelength IR radiation, allows for the direct integration of this method with conventional optical spectroscopy and microscopy applied in biological studies and materials research. Furthermore, higher spatial resolutions can be obtained with visible light (400-700 nm) as the diffraction-limit is ca. 20-times sharper than for LWIR (7-14 μm); this potentially extends the utility of remote thermography to intracellular, *in vitro*, and *in vivo* studies[17].

To promote the advancement and widespread use of remote thermography, a much broader portfolio of luminescent, thermally sensitive, and temperature-range tunable materials is required. These emitters must exhibit a fully reproducible radiative lifetime *vs.* temperature dependence, and demonstrate an invariant behavior towards excitation-light intensity. While emitters satisfying these conditions do exist, the precision of thermometry utilizing them was so far reported to be only in the 0.1-1 °C range[9,15,18]. This level of precision can be again attributed to two factors: the often moderate thermal sensitivity of available thermographic luminophores[18], and the PL lifetime measurement techniques that have been traditionally applied in industry. As a result, progress in the development of PL-lifetime thermography suffered from the exclusive use of expensive, bulky techniques that are in fact single-point measurements, which prohibit fast image acquisition. To address these shortcomings, we present (i) a new family of low-dimensional tin-halide luminophores well-suited for remote thermography due to their strongly temperature-dependent, compound-specific PL lifetimes, (ii) a high thermographic precision down to 0.05 °C with operation in a broad temperature range from -100 to 110 °C, and (iii) a new thermographic method utilizing ToF cameras[19,20] for cost-effective, high-resolution, fast thermal imaging.

**Results**

**Low-dimensional tin-halide perovskite-derived luminophores.** In the search for suitable thermographic luminophores, we analyzed the potential of lead- or tin halides with perovskite or lower-dimensional perovskite-like crystal structures. These compounds are formed by metal-halide polyhedral anions, typically $MX_6^{4-}$ octahedra (M = Pb, Sn), which are either fully isolated and surrounded by positively-charged counterions (so-called 0D-compounds) or connected into extended one-, two- or three-dimensional (1D-3D) frameworks through corner-, edge- or face-sharing[21,22]. In particular, 3D lead-halide perovskites, have recently emerged as prominent optoelectronic materials for photovoltaics and photodetectors[23-27], hard-radiation detection[28-30], as well as bright light emitters[31-34]. Although bright luminescence has been reported at all dimensionalities for metal-halides, we have found that only 0D- and 1D-compounds exhibit a suitable set of optical characteristics for remote thermography.

In low-dimensional, and specifically zero-dimensional perovskites, the electronic structure evolves from disperse electronic bands in 2D-3D materials to more localized states (molecule-like) in 0D-1D compounds[21]. Stemming from this, the mechanism of luminescence is also drastically different, and ranges from rather large, delocalized Wannier-type excitons in 3D materials to ultra-small Frenkel-like self-trapped excitons (STEs) in the 0D-compounds[35-38]. We have tested a range of previously reported and new highly-luminescent, fully-inorganic and hybrid organic-inorganic 0D and 1D tin-halide compounds for their suitability for thermography in the range of -40 to 120 °C. Based on temperature-dependent PL and PL-lifetime measurements, three suitable candidates have been shortlisted: $[C(NH_2)_3]_2SnBr_4$ $[C(NH_2)_3$ = guanidinium; CCDC code 1854819], $Cs_4SnBr_6$[39] and $(C_4N_2H_{14}I)_4SnI_6$[40] (Fig. 2; Table 1). All of these compounds exhibit temperature-dependent PL lifetimes in the range of 1 ns – 1 μs, and thus ideally match the optimal modulation range for most commercial ToF sensors (e.g. from tens of kHz to tens of MHz). This is in sharp contrast to the thermographic luminophores based on rare-earth-doped oxides[9,13,16,41,42], which are characterized by much slower emission that is typically in the ms-range. Furthermore, the absorption coefficients of tin halides for the UV-A range (315–400 nm, convenient for optical excitation) are high, and they are comparable to the bandgap absorption values (Fig. 2) as each octahedron within the structure can act as an absorber/emitter. In contrast, the UV-A range absorption of rare-earth-doped oxide luminophores is weak due to a limited concentration of dopants (activator) or absorbing centers[9].

Both solid-state and solution-based methods were used to prepare the three types of luminophores utilized within this work. The fully-inorganic, 0D $Cs_4SnBr_6$ was synthesized through a solid-state approach in which a mixture of CsBr and $SnBr_2$ where repeatedly pressed and heated at sub-melting temperatures as described in our recent report[39]. $Cs_4SnBr_6$ (*R-3c* space group) is composed of $[SnBr_6]^{4-}$ octahedra separated by $Cs^+$ cations (Fig. 2a)[43]. A new 1D hybrid organic-inorganic compound, guanidinium tin-bromide, $[C(NH_2)_3]_2SnBr_4$ (*Pna2₁* space group), was crystallized from a solution of the respective ions in concentrated HBr. Its tin-halide backbone consists of corner-sharing $[SnBr_5]_\infty^{2-}$ square pyramids (Fig. 2b). Another 0D hybrid compound, $(C_4N_2H_{14}I)_4SnI_6$, was prepared by co-precipitation from a dimethylformamide solution according to Ref.[40] $(C_4N_2H_{14}I)_4SnI_6$ comprises isolated $[SnI_6]^{4-}$ octahedra surrounded by large, organic cations (Fig. 2c). Powder X-ray diffraction patterns of all three materials indicate high phase-purity (Supplementary Figs. 1-3).

**Optical properties of low-dimensional tin-halides.** Although diverse in their compositions, structures and syntheses, all present materials are qualitatively unified by their broadband and highly Stokes-shifted PL. Their optical absorption behavior was determined through the measurement of both PL excitation (PLE) and absorption via the application of the Kubelka-Munk (K-M) transformation to diffuse reflectance spectra (Fig. 2d-f). The absorption of 0D tin-halides, $Cs_4SnBr_6$ and $(C_4N_2H_{14}I)_4SnI_6$, appears as molecular-like bands that coincide with the PLE spectra. Additionally, both materials exhibit absorption features at shorter wavelengths that do not contribute to emission (Fig. 2d,f). This differs from the 1D compound, $[C(NH_2)_3]_2SnBr_4$, which instead exhibits a continuous PLE spectrum (Fig. 2e). Such behavior can be associated with the partial band dispersion that occurs along the polyhedral chain in 1D metal-halides[38].

These molecular-like characteristics of the absorption spectra of low-dimensional compounds are reflected in their emission. Unlike CsSnBr$_3$ (a 3D perovskite semiconductor) that exhibits weakly Stokes shifted and narrow PL (as a result of excitonic-recombination) with very low quantum yield (QY)[44,45], Cs$_4$SnBr$_6$ shows room-temperature (RT), green broadband PL centered at 535 nm, with a full-width at half-maximum (FWHM) of 120 nm, and a QY of about 20% (Fig. 2b, Table 1). Such broadband and highly Stokes shifted emission has been observed for other low-dimensional metal-halide compounds such as [C(NH$_2$)$_3$]$_2$SnBr$_4$, (C$_4$N$_2$H$_{14}$Br)$_4$SnBr$_6$, and (C$_4$N$_2$H$_{14}$I)$_4$SnI$_6$, and is commonly associated with STE recombination (Fig. 2b,d,f and Supplementary Fig. 4)[35-38].

The materials typically used in remote thermometry are either rare-earth-doped phosphors exhibiting emission from charge-transfer states[9] or transition metal oxides with STE based emission[46]. As a result of thermal de-trapping, the emission from thermographic luminophores is generally mono-exponential and strongly dependent on the material's temperature; the typical relaxation time for this process is in the range of µs to ms. With the low-dimensional tin-halides, presented here, we have also found the emission lifetime strongly temperature-sensitive, and we associate this with STE de-trapping. The corresponding energy diagram in Fig. 3a depicts this process: upon photon absorption (1), an electron is promoted to an excited state and, after its thermalization (2), is trapped (3) in a long-lived STE state. This trapping is then followed by a radiative recombination with broadband emission (4). A thermally assisted de-trapping pathway (5), followed by fast non-radiative recombination, is also present and plays a key role in the temperature-dependence of the PL characteristics. During de-trapping, a distorted lattice around an STE can be returned back to its original state through exciton-phonon coupling. Thus higher temperatures facilitate de-trapping and assist relaxation via a fast non-radiative channel. In agreement with this mechanism, a strong thermally-driven acceleration of the PL lifetime is observed (measured with time-resolved PL, TRPL, Fig. 3b). Furthermore, the PL intensity scales with the change in PL lifetime (Supplementary Fig. 5), and its change is fully reversible with temperature (Supplementary Fig. 6). The absolute PL QYs within the thermal sensitivity range vary linearly between *ca*. 100 % at lower temperatures to less than 1% at higher temperatures. The highly useful, practical assets of this STE-based emission are (i) the independence of PL lifetime from excitation intensity (Supplementary Fig. 7) and (ii) the fact that the PL lifetime is the same throughout the whole emission band (Supplementary Fig. 8). In addition, the PL lifetime is also independent of the material's environment (freestanding or encapsulated within a polymer matrix, Supplementary Fig. 9), and it is highly reproducible from batch-to-batch (Supplementary Fig. 10) despite differences in the degree of crystallinity and purity. This suggests that the origin of the temperature-dependence is not related to the freezing-out of defect or trap states, but rather a phonon-assisted de-trapping process that is followed by fast non-radiative relaxation. This set of PL characteristics is clearly advantageous over semiconductive metal-halides with delocalized electronic structures (2D, 3D-compounds), wherein the PL decay is a complex function of the density, types, and depths of defect states, as well as the electronic doping level, the state of the surface, and size-quantization. As a result, there is often high batch-to-batch variability; even within similar synthetic methods.

Such behavior of the STE emission, in terms of PL lifetime and PL QY, is not unique to $Cs_4SnBr_6$, but is also shared by the other tested 0D and 1D metal-halides (Fig. 3c). Due to the phonon-assisted thermal effect, the temperature range for thermal sensing can be compositionally and/or structurally engineered to either higher or lower temperatures than in the case of $Cs_4SnBr_6$. In the temperature range from -200 °C to 110 °C, $[C(NH_2)_3]_2SnBr_4$ begins to exhibit PL lifetime acceleration at the lowest temperatures, with a sensitivity range of -100 to -30 °C; $Cs_4SnBr_6$ has a sensitivity range from -30 °C to 40 °C, whereas $(C_4N_2H_{14}I)_4SnI_6$ exhibits sensitivity from 40 °C up to 110 °C (these ranges are shown as colored areas in Fig. 3c). Surprisingly, $(C_4N_2H_{14}Br)_4SnBr_6$, also exhibits broadband STE emission, but our measurement setup was not able to reach the thermally sensitive range. This suggests that this range lies at much higher temperatures (Supplementary Fig. 11), and the use of this material as a thermographic luminophore would be rather limited by its thermal stability.

Highly reproducible variation of the PL lifetime, by several orders of magnitude (*e.g.* 2 orders over a 100 °C range for $Cs_4SnBr_6$), makes these metal-halides potent luminophores with high thermometric precision. To assess the practically achievable resolution for thermographic applications, we considered the uncertainty in the monoexponential fitting for $Cs_4SnBr_6$ near RT (<1 ns, or *ca.* 0.2 % of the absolute lifetime value) and applied this error to each point of the PL lifetime *vs.* temperature graph (Fig. 3d, inset). Remarkably, these experimentally determined lifetimes follow a near-linear trend with a similar uncertainty of 1 ns that validates the above estimation. In the vicinity of RT, the lifetime *vs.* temperature dependence for $Cs_4SnBr_6$ gives a sensitivity of about 17 ns °C$^{-1}$ estimated as $\frac{d\tau}{dT}$ (inset to Fig. 3d). This then yields a thermometric precision of 0.05 °C, which is several times better than previous estimates for fluorescent lifetime thermography[18]. Furthermore, an additional figure-of-merit for thermographic luminophores is the specific sensitivity, $\alpha = \frac{1}{\tau}\frac{d\tau}{dT}$, which in the case of low-dimensional tin-halides reaches values of 0.06 °C$^{-1}$ (Supplementary Fig. 12). This is among the highest reported values for thermographic luminophores[47]. Higher resolutions have been demonstrated, but only for a specific case where operation was limited to a narrow temperature range around a phase transition.[47] Further work is needed to shed light into the physics of the STE emission of these novel luminophores. Herein, we have applied several models to fit the experimental temperature-dependencies of PL lifetime (Supplementary Note 1; Supplementary Figs. 13-15; Supplementary Tables S1-3). Although they all provide a satisfactory fit, each has difficulties providing physically meaningful fitting parameters. For instance, the Mott model, commonly used for thermal luminophores[8], results in rather unphysical values for the activation temperature, whereas the exciton-phonon scattering model[48] suggests the participation of up to 13 phonons per single de-trapping event.

**Time-of-flight thermography using low-dimensional tin-halides.** Although several time-resolved measurement techniques like PL decay trace-measurements, TCSPC (Fig. 3b-d) or phase fluorometry (frequency domain time-resolved fluorescence) could be used to precisely measure the PL-lifetime of these materials, all of these have traditionally been limited by the fact that they only use a single channel detector[49]. Consequently, lengthy acquisition through point-by-point scanning is required, and this severely limits the thermographic image capture rate.

As an innovative solution to this problem, we adopted the use of ToF-FLI[19,20] – a frequency domain time-resolved technique that can be used to acquire a 2D map of PL-lifetimes[50] – and combined it with thermally sensitive luminophores. ToF-FLI is thus for the first time used for thermographic imaging. This approach offers both the rapid acquisition speed and the excellent depth precision of ToF detectors such as those found in consumer electronics (*e.g.* the Kinect 2.0), which we confirmed to be several mm at a distance of about 2 m (inset to Fig.4a; Supplementary Fig. 16)[51]. By converting this depth variation into an equivalent delay time (details in Supplementary Note 2 and Supplementary Fig. 17), we find that such a technique could have a precision in the range of tens-to-hundreds of picoseconds, and this suggested that lifetime precisions approaching those of conventional TCSPC methods were possible. Furthermore, this level of precision is achieved in real-time with video recording. Briefly, the working principle for such a ToF-FLI image sensor is based on the acquisition of four, phase-locked images at 0°, 90°, 180° and 270° phase differences with respect to the excitation signal ($I_0$, $I_1$, $I_2$ and $I_3$ on Fig. 4b) followed by the recalculation of the average intensity $I$, modulation depth $M$ and phase delay, $\Delta\phi$ (details described in Supplementary Note 3).

To demonstrate the concept of affordable ToF-FLI thermography with low-dimensional tin-halides, we used a compact, stand-alone prototype developed by some of the co-authors from CSEM (Switzerland), in which all the necessary hardware components for wide-field frequency-domain FLI were incorporated (Supplementary Fig. 18, Supplementary Note 4). As a test of the thermographic performance of our system, we deposited a $Cs_4SnBr_6$ powder over a resistively heated pattern and enclosed the material with a glass coverslip, and then measured the resulting lifetime image of the heated pattern. Compared to a conventional bolometric thermogram (Fig. 4c,d), the ToF-FLI method showed a much higher lateral thermographic resolution. The blurring observed in the bolometric thermogram of the sample is mostly due to absorption by the coverslip in the mid- to far-IR range (compare Fig. 4c with the bolometric thermogram of the uncovered ITO pattern during heating in Supplementary Fig. 19). These absorption effects highlight a major obstacle, *i.e.* absorption of thermal emission by glass and similar media, which prevent the combination of bolometric thermography with conventional optical systems that utilize glass or quartz elements.

While a still image demonstrates the ability of this system to precisely measure a 2D map of PL-lifetime, observing dynamic processes requires the ability to record video and to measure at sufficiently high rates. To demonstrate the potential of our ToF-FLI prototype – whose sensor can record with a rate of up to 100 frames per second – for thermographic video acquisition, we recorded a video of the thermal response of a $(C_4N_2H_{14}I)_4SnI_6$ powder through 1 mm of a glass substrate to the brief contact of a soldering pin (temperature at the apex was approx. 120 °C; ToF-FLI specifications in Supplementary Note 4; Supplementary Video 1). Indeed, it was possible to observe the dynamic temperature changes that occurred as well as the heat transfer through and along the substrate - a challenge for pixel-by-pixel scanning technologies.

In summary, we discovered that the de-trapping process of STEs in low-dimensional tin-halides exhibits extreme thermal sensitivity over a compositionally tunable range of temperatures. In particular, such emission is characterized by monoexponential decays with a steep dependence of PL lifetime on temperature (up to 20 ns °C$^{-1}$). We then applied these features to high-precision thermometric measurements over a wide temperature range (-100 °C to 110 °C), and furthermore demonstrated a novel approach to remote optical thermography by combining these low-dimensional tin-halide luminophores with ToF-FLI. By doing so, we have succeeded in achieving low-cost, precise, and high-speed PL-lifetime thermographic imaging.

**Methods**

**Synthesis of Cs$_4$SnBr$_6$.** All chemicals were used as received without further purification. All manipulations were performed air-free inside a glovebox with $H_2O$ and $O_2$ levels <0.1 ppm. CsBr (99 %, ABCR; 99 %, Alfa Aesar) and SnBr$_2$ (99.2 %, Alfa Aesar) were mixed in a 4.5:1 molar ratio, mortared, and pressed together into a pellet (> 5 tons of pressure, 13 mm die). The pellet was then sealed under vacuum ($10^{-2}$ – $10^{-3}$ mbar) in a pyrex tube and heated to 350 °C for 60 hours. The ampule was opened in the glovebox, and the above process was repeated once more. The pseudo-binary CsBr-SnBr$_2$ phase diagram[43] indicates that temperatures above 380 °C result in a peritectic decomposition.

**Synthesis of [C(NH$_2$)$_3$]$_2$SnBr$_4$.** [C(NH$_2$)$_3$]$_2$SnBr$_4$ was crystallized from hydrobromic acid (HBr, 48 % water solution, Acros) under inert conditions (Ar atmosphere) in a 20-mL Schlenk vessel. For this, Sn powder (0.250 g, 2.1 mmol, 99.8 %, ~325 mesh, from Acros) was dissolved in 3 mL of HBr (degassed under stirring in Ar atmosphere for ~20 min. beforehand). First, the mixture was stirred for 10 min. at RT and then heated with a glycerol bath at 80 °C. When all Sn had dissolved, 0.378 g of guanidinium carbonate, [C(NH$_2$)$_3$]$_2$CO$_3$, (99+ %, Acros) was carefully added; a strong evolution of gas was observed. The reaction mixture was then stirred for an additional 5 min. at 80 °C resulting in a clear colorless solution, followed by nature cooling. In 8-10 hr., a white crystalline powder of [C(NH$_2$)$_3$]$_2$SnBr$_4$ in the shape of thin needles was separated by vacuum filtration under Ar flow and dried under vacuum.

**Synthesis of (C$_4$N$_2$H$_{14}$I)$_4$SnX$_6$ (X = Br, I).** C$_4$N$_2$H$_{14}$X$_2$ was prepared according to Ref.[40] with small modifications. 7.4 mL (0.056 mol) of hydroiodic acid (HI, 57% water solution, ABCR) or 6.4 mL (0.056 mol) hydrobromic acid (HBr, 48% water solution, Acros) was added to 3 mL (0.028 mol) of N,N'-dimethylethylenediamine in 20 mL of ethanol at 0 °C and stirred overnight. A white-yellowish powder of C$_4$N$_2$H$_{14}$X$_2$ was obtained after removing the solvent under vacuum. The salt was washed several times with diethyl ether and dried under vacuum. C$_4$N$_2$H$_{14}$X$_2$ was stored in the glovebox for future use. A precursor solution was prepared by mixing 0.1 mmol of SnX$_2$ and 0.4 mmol of C$_4$N$_2$H$_{14}$X$_2$ in 1 mL of degassed DMF overnight. 0.5 mL of the precursor solution was then rapidly injected, under stirring, into either 3 mL of anhydrous toluene with 30 mL of trioctylphosphine (for X = I) or into 3 mL anhydrous toluene (X = Br). This was followed by the immediate formation of the precipitate, (C$_4$N$_2$H$_{14}$I)$_4$SnX$_6$. The mixture was stirred at RT for another 15 min, the crude solution was centrifuged and the supernatant was discarded. The precipitate was washed with anhydrous toluene two more times, followed by drying under vacuum. The whole procedure was performed air-free and $(C_4N_2H_{14}I)_4SnX_6$ was stored in the glovebox.

**Optical characterization**

**UV-Vis absorbance spectra** were obtained using the Kubelka-Munk transformation of diffuse reflectance of the microcrystalline powders that were collected using a Jasco V670 spectrophotometer equipped with a deuterium ($D_2$) lamp (190 – 350 nm) for UV, a halogen lamp (330 – 2700 nm) for UV/NIR, and an integrating sphere (ILN-725) with a working wavelength range of 220 – 2200 nm.

**Photoluminescence emission**, **excitation and quantum yield.** PL and PLE spectra were measured with a Fluorolog iHR 320 Horiba Jobin Yvon spectrofluorometer equipped with a Xe lamp and a PMT detector. Absolute values of PL QY were measured using a Quantaurus-QY spectrometer from Hamamatsu in powder mode.

**Time-resolved and steady-state photoluminescence temperature dependence.** Samples were located on top of a 4-stage Peltier cooling/heating element in an evacuated chamber with a quartz window. The sample temperature was adjusted and stabilized using a self-made electronic scheme based on an Arduino microcontroller and thermocouple sensor. The current through the Peltier was reversible; hence the setup provided a wide working temperature range of -40 ℃ to 120 ℃. This is an open-source project by the authors, which is deposited and described in detail at https://www.researchgate.net/project/High-power-thermoelectric-cooler-TEC-controller-with-4-stage-Peltier-refrigerator-heater. Measurements in the low temperature range (78-300 K) were performed using a Joule–Thomson cryostat (MMR Technologies) with a cooling/heating rate of about 2-5 K/min. TRPL traces were recorded with a heating rate of 2 ºC /min. A 355 nm excitation source (a frequency-tripled, picosecond Nd:YAG laser, Duetto from Time-Bandwidth) and a CW diode laser with an excitation wavelength of 405 nm (for steady-state PL) were used. Scattered emission from the lasers was filtered out using dielectric long-pass filters with cut-offs at 400 nm and 450 nm, respectively. For TRPL, the signal was acquired using a time-correlated single photon counting (TCSPC) setup, equipped with a SPC-130-EM counting module (Becker & Hickl GmbH) and an IDQ-ID-100-20-ULN avalanche photodiode (Quantique) for recording the decay traces.

**ToF thermography imaging.** As a heating element we used a patterned ITO thin film deposited on a microscopy cover slide (Structure Probe, Inc. USA). The patterning was performed by a standard "wet" lithography process using a positive photoresist and a contact mask that was ink-jet printed onto a transparent polymer film. The heat distribution in the obtained ITO pattern was checked by a LWIR camera: Thermal Seek customized with ZnSe lensed macro-objective made according to the project: https://www.thingiverse.com/thing:525605. $Cs_4SnBr_6$ powder was then dispersed on top of the ITO pattern and encapsulated with a photopolymer glue (ZLD-312 UV, Zhanlida Co., Ltd.) and a second glass microscopy cover slide under inert atmosphere in a nitrogen-filled glovebox.

The ToF-FLI prototype includes a modulated light source (LED, 365 nm emission wavelength), a CMOS ToF imager (256 x 256 pixels) originally developed by CSEM for 3D imaging, dedicated FPGA-based electronics, and optical components for illuminating the probe and collecting the PL emission. The camera electronics are based on a stacked PCB approach, including a base board, a FPGA processing module and a sensor head PCB. A MATLAB GUI running on a separate PC is used to set the measurement parameters (such as modulation frequency, illumination intensity and integration time) and display the results. The LED modulation frequency can be varied between 3 kHz and 20 MHz, allowing the measurement of PL lifetimes from hundreds of microseconds down to a few nanoseconds with sub-nanosecond precision. With the current optics an area of 5.3x5.3 mm is imaged. The emission wavelength can be selected by exchangeable spectral filters. More detailed technical characteristics of the ToF-FLI image sensor setup are listed in Supplementary Note 3 and Supplementary Reference 4.

## Data availability

The data that support the findings of this study are available from the corresponding authors upon reasonable request.

## Acknowledgements

M.K. acknowledges financial support from the European Union through the FP7 (ERC Starting Grant NANOSOLID, GA No. 306733). C.H. and S.C. thank the Swiss Nano-Tera program (projects FlusiTex and FlusiTex Gateway) and the Swiss Commission for Technology and Innovation CTI (project SecureFLIM) for financing the development of the ToF-FLI imager. Authors thank Gabriele Rainò and Stefan T. Ochsenbein for fruitful discussion.

## Author contributions

This work originated from continuing interactions between the research groups at ETH Zurich and CSEM. S.Y., B.B., Y.S. performed measurements; C.H., and S.C. developed and adapted the FLI reader; B.B., O.N., D.D. and M.B. synthesized the tin-halide thermographic luminophores; S.Y. and Y.S. analyzed the results; S.Y., B.B., and M.K. wrote the manuscript. M.K. supervised the work. S.Y. and B.B. contributed equally to this work. All authors discussed the results and commented on the manuscript.

**Additional Information**

**Competing financial interests:** The authors declare no competing financial interest.

**Reprints and permission** information is available online at http://npg.nature.com/reprintsandpermissions

**Table 1. Structural and optical characteristics of tin-halide thermographic luminophores.**

| Composition | Space group | Structure | Absorpt. max. | | Emiss. max. | | Emission FWHM | | Stokes shift | QY (@RT) | Temp. range |
|---|---|---|---|---|---|---|---|---|---|---|---|
| | | | nm | eV | nm | eV | nm | eV | eV | % | °C |
| [C(NH₂)₃]₂SnBr₄ | *Pna2₁* | 1D | 350 | **3.55** | 555 | **2.24** | 125 | **0.5** | **1.31** | 2 | [-100, -30] |
| Cs₄SnBr₆ | *R-3c* | 0D | 345 | **3.6** | 535 | **2.32** | 120 | **0.51** | **1.28** | 20 | [-30, 40] |
| (C₄N₂H₁₄I)₄SnI₆ | *P-1(#2)* | 0D | 400 | **3.11** | 630 | **2.0** | 125 | **0.4** | **1.14** | 75 | [40, 110] |
| (C₄N₂H₁₄Br)₄SnBr₆ | *P-1(#2)* | 0D | 335 | **3.71** | 575 | **2.18** | 107 | **0.4** | **1.54** | ~100 | N/A |

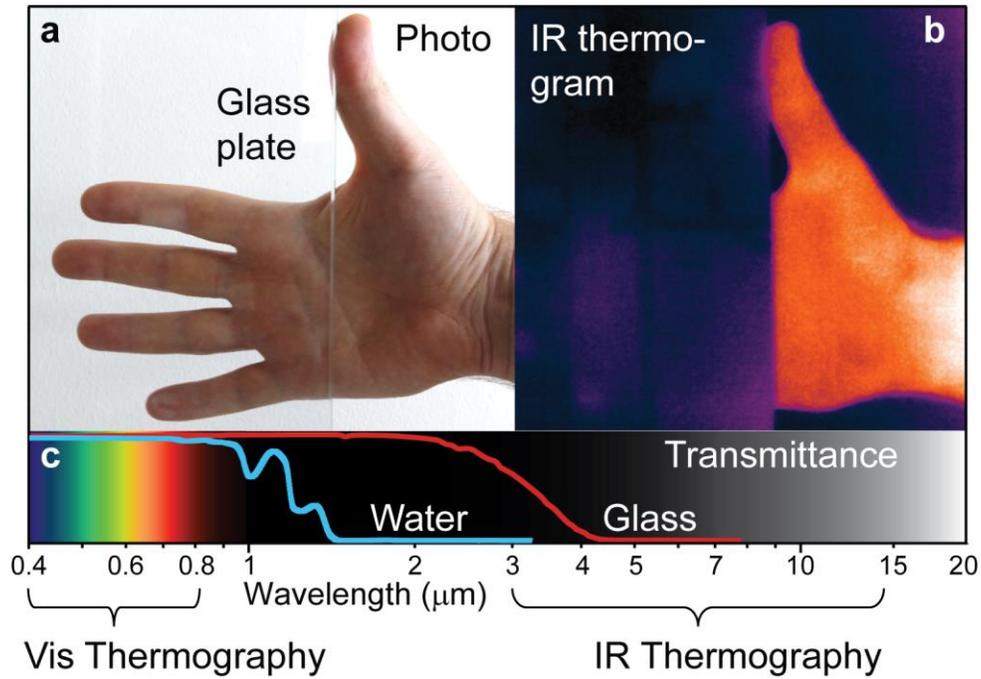

**Figure 1. Visible-light and infrared (IR) thermography comparison.** Differences in transmittance for visible and IR light are demonstrated in (a) photograph, and in (b) IR thermogram captured using a commercial Seek Thermal Compact Pro™ LWIR bolometry camera. (c) The transparency ranges of two ubiquitous optical media.

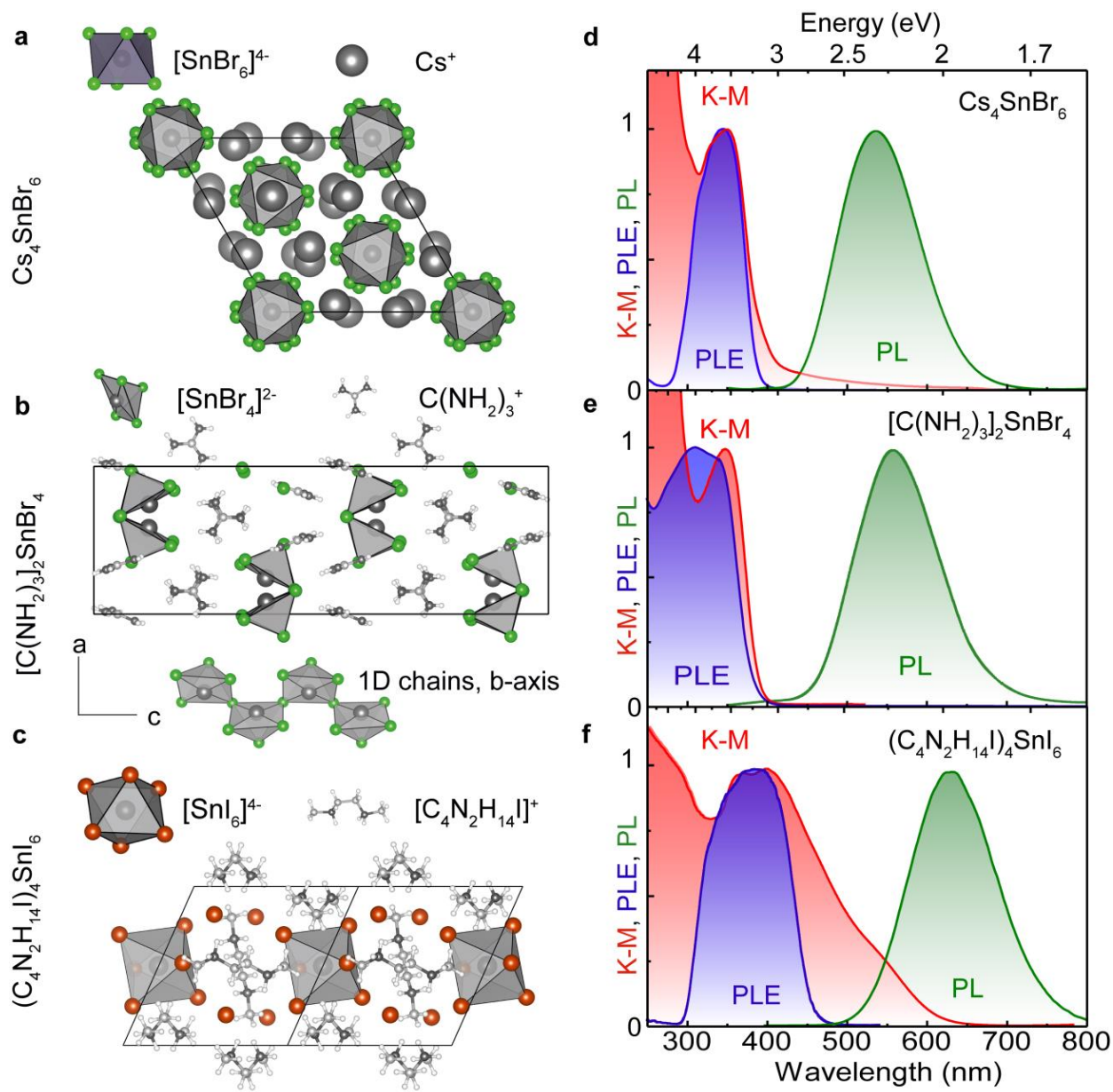

**Figure 2. Crystallographic structures and basic optical properties of select thermographic luminophores.** Crystal structure of (a) $Cs_4SnBr_6$, (b) $[C(NH_2)_3]_2SnBr_4$ and (c) $(C_4N_2H_{14}I)_4SnI_6$ and their corresponding absorption (Kubelka-Munk transformed spectrum), PL, and PLE spectra.

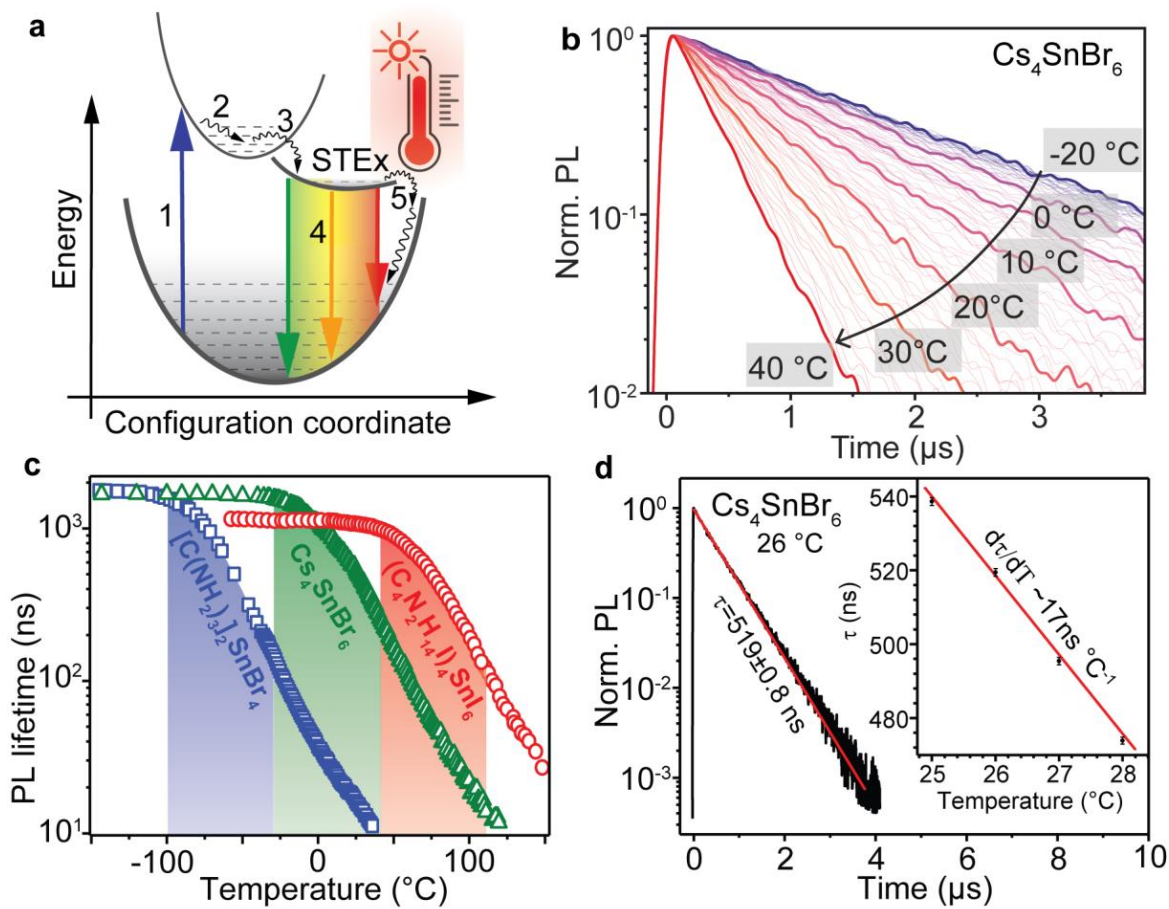

**Figure 3. Thermal effects on PL lifetime variation of self-trapped excitonic (STE) emission in low-dimensional tin-halides.** (a) Energy diagram depicting the STE processes: 1 - photon absorption, 2 - thermalization, 3 - trapping, 4 - radiative recombination, 5 -thermally-assisted detrapping followed by non-radiative recombination. (b) Temperature evolution of TRPL traces for $Cs_4SnBr_6$ excited at 355 nm. (c) PL lifetime temperature dependence for $[C(NH_2)_3]_2SnBr_4$ (blue curve), $Cs_4SnBr_6$ (green curve), $(C_4N_2H_{14}I)_4SnI_6$ (red curve). (d) The monoexponential model demonstrates a high precision in lifetime fitting of about 1 ns. The inset to Fig.3d illustrates the accuracy of PL lifetime *vs.* temperature dependency within error bars of 1 ns that correspond to a precision of 0.05 °C.

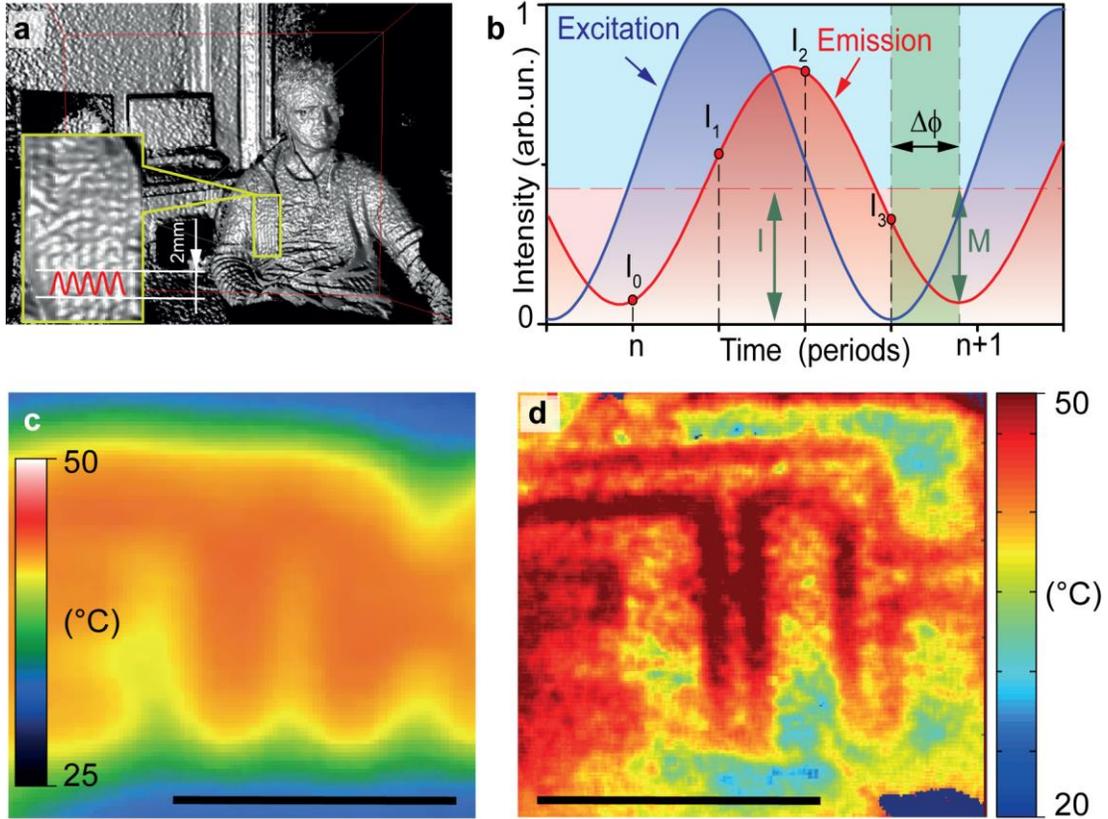

**Figure 4. Demonstration of the principles for remote thermography based on ToF sensors.** (a) 3D depth image acquired with the ToF camera of a Kinect 2.0™ showing a depth resolution of 2 mm in the inset. This is equivalent to a precision of about 10 ps in the time-domain. (b) Recalculation of phase-locked intensities to a phase-shift in the frequency domain. (c) Thermographic image of a sample that consists of encapsulated $Cs_4SnBr_6$ powder placed between patterned ITO and a glass coverslip, and then acquired with a commercial Seek Thermal Compact Pro™ LWIR bolometry camera equipped with a ZnSe lensed macro-objective. The image was taken as a current passed through the ITO resulting in resistive heating. (d) ToF-FLI thermogram of the same sample under the same heating conditions from (c). Scale bars in (c) and (d) are 3 mm.

# Supplementary Information

**High-resolution remote thermography using luminescent low-dimensional tin-halide perovskites**

*Sergii Yakunin, [1,2#*] Bogdan M. Benin, [1,2#] Yevhen Shynkarenko, [1,2] Olga Nazarenko, [1,2] Maryna I. Bodnarchuk,[1,2] Dmitry N. Dirin,[1, 2] Christoph Hofer,[3] Stefano Cattaneo[3] and Maksym V. Kovalenko[1,2*]*

[1] Laboratory of Inorganic Chemistry, Department of Chemistry and Applied Biosciences, ETH Zürich, CH-8093 Zürich, Switzerland

[2] Laboratory for Thin Films and Photovoltaics, Empa – Swiss Federal Laboratories for Materials Science and Technology, CH-8600 Dübendorf, Switzerland

[3] Swiss Center for Electronics and Microtechnology (CSEM), Center Landquart, CH-7302 Landquart, Switzerland

# Equal contributors. * Corresponding authors.
[*] E-mails:  mvkovalenko@ethz.ch; yakunins@ethz.ch

Table of Contents

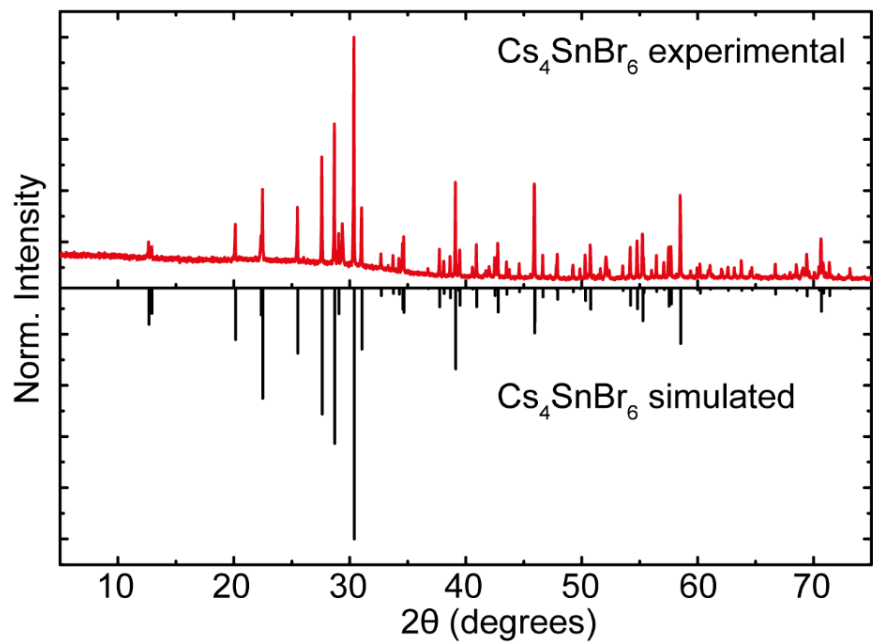

**Supplementary Figure 1.** Powder X-ray diffraction pattern of Cs$_4$SnBr$_6$.

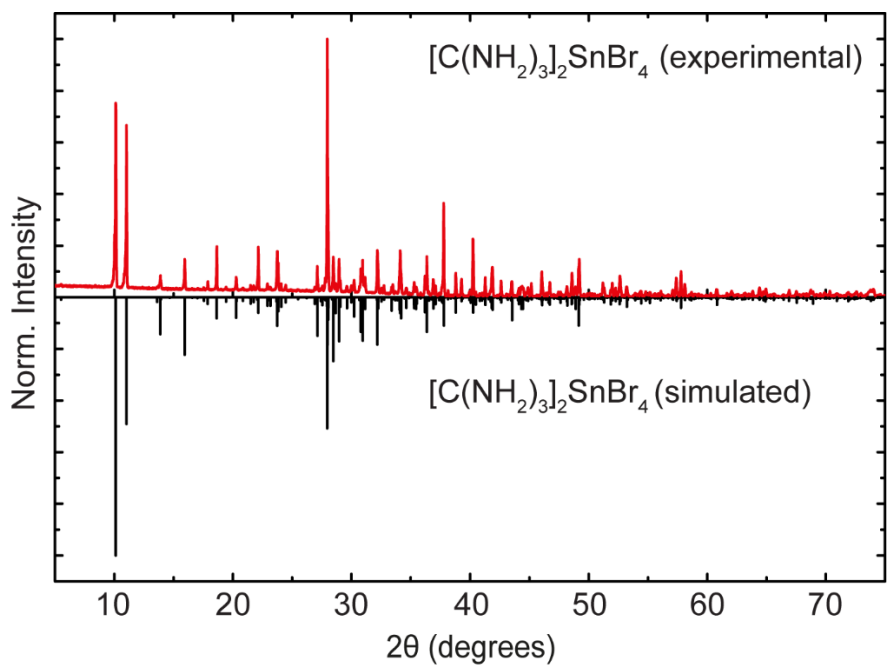

**Supplementary Figure 2.** Powder X-ray diffraction pattern of [C(NH$_2$)$_3$]$_2$SnBr$_4$.

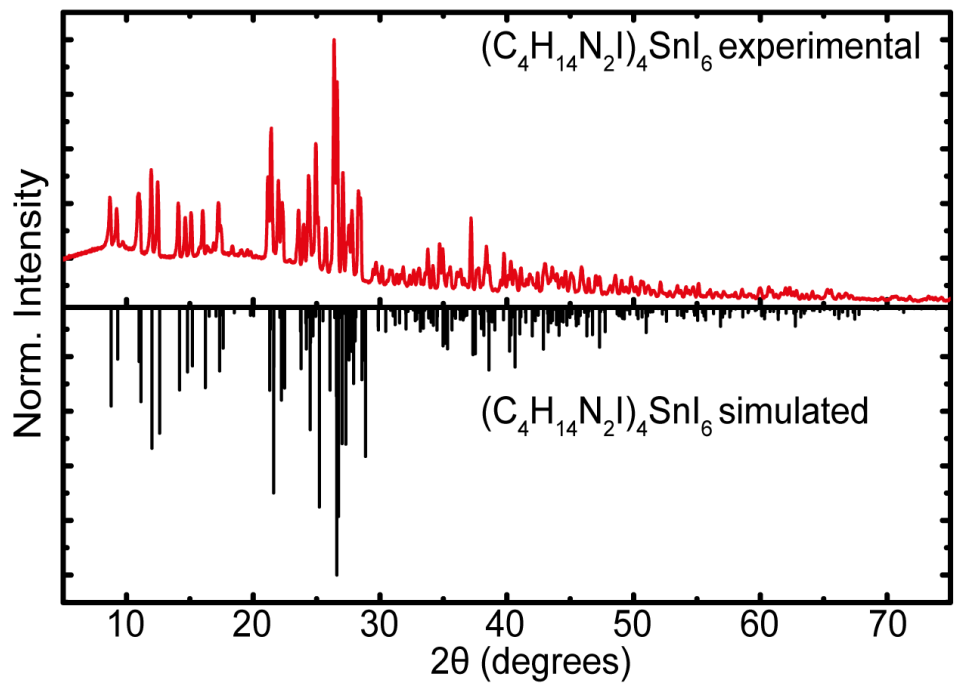

**Supplementary Figure 3.** Powder X-ray diffraction pattern of $(C_4H_{14}N_2I)_4SnI_6$.

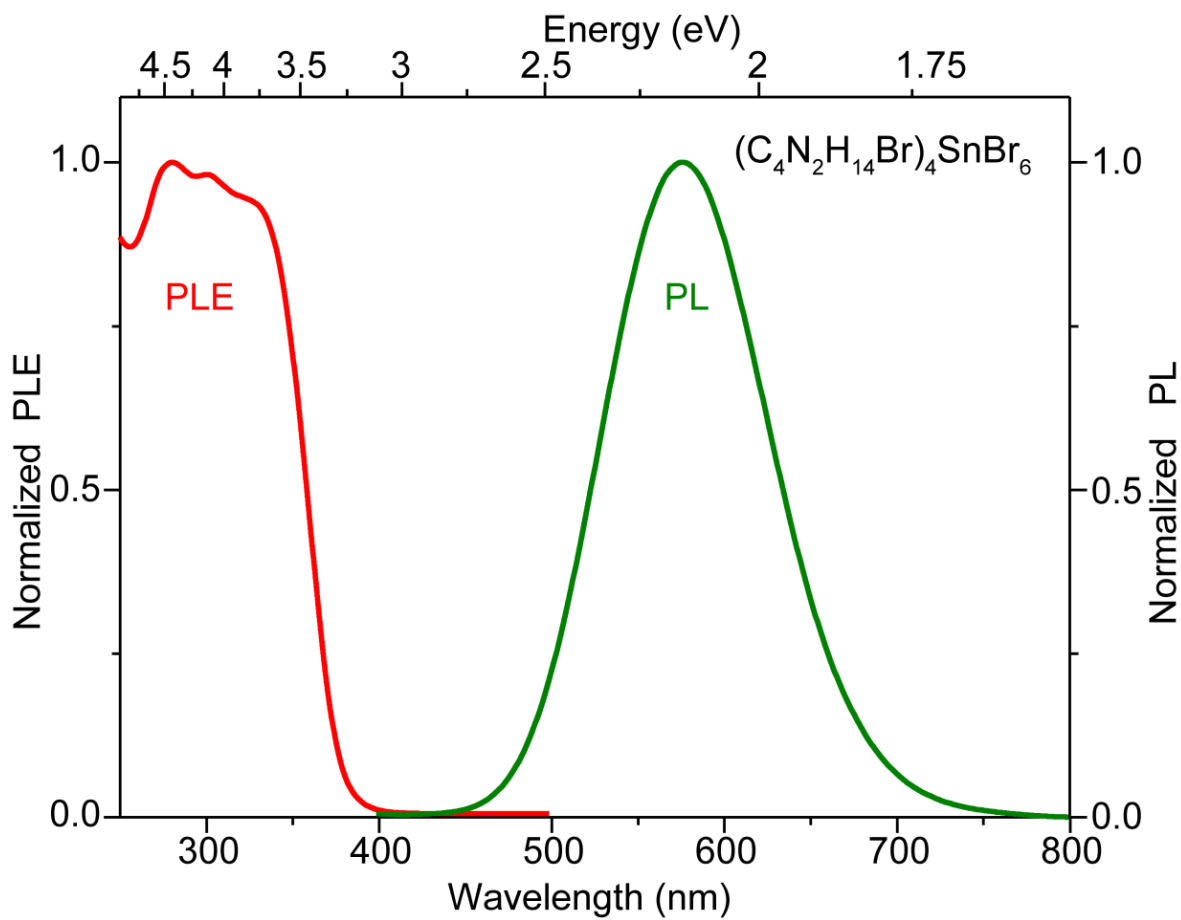

**Supplementary Figure 4.** Optical characterization of (C$_4$N$_2$H$_{14}$Br)$_4$SnBr$_6$: photoluminescence excitation (PLE, red) and photoluminescence (PL, green) spectra.

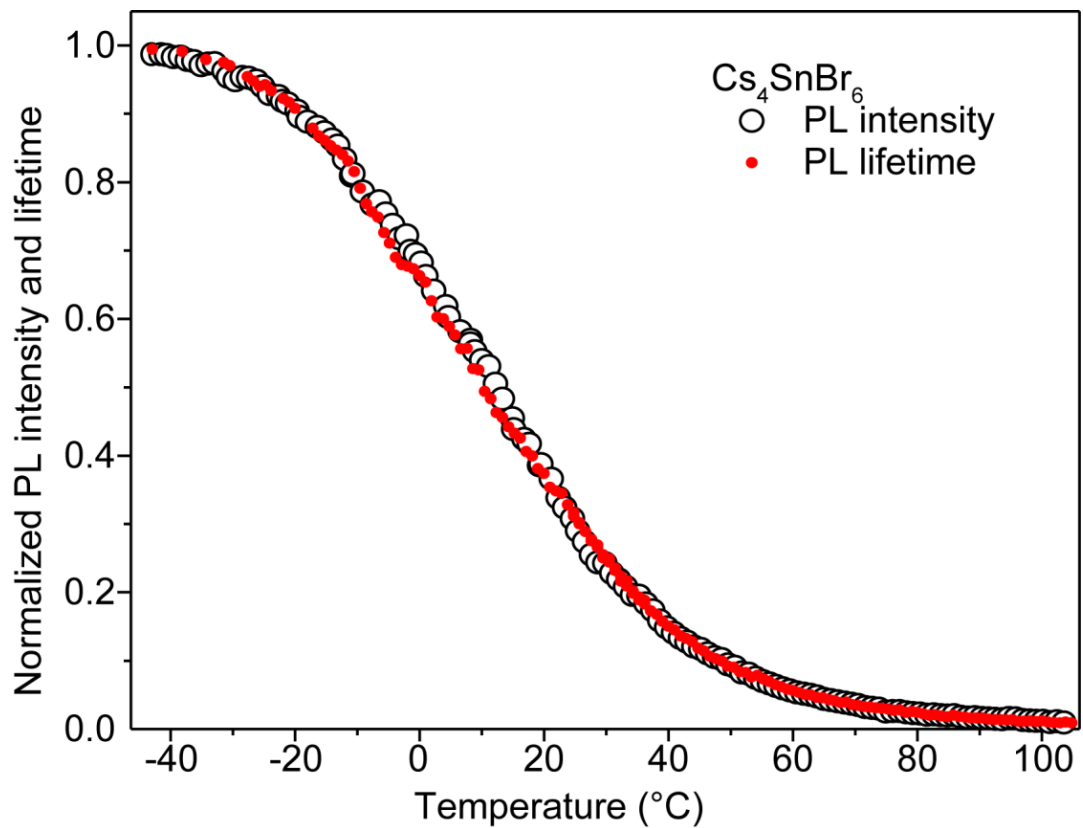

**Supplementary Figure 5.** Temperature dependence of Cs$_4$SnBr$_6$ PL emission lifetime and intensity.

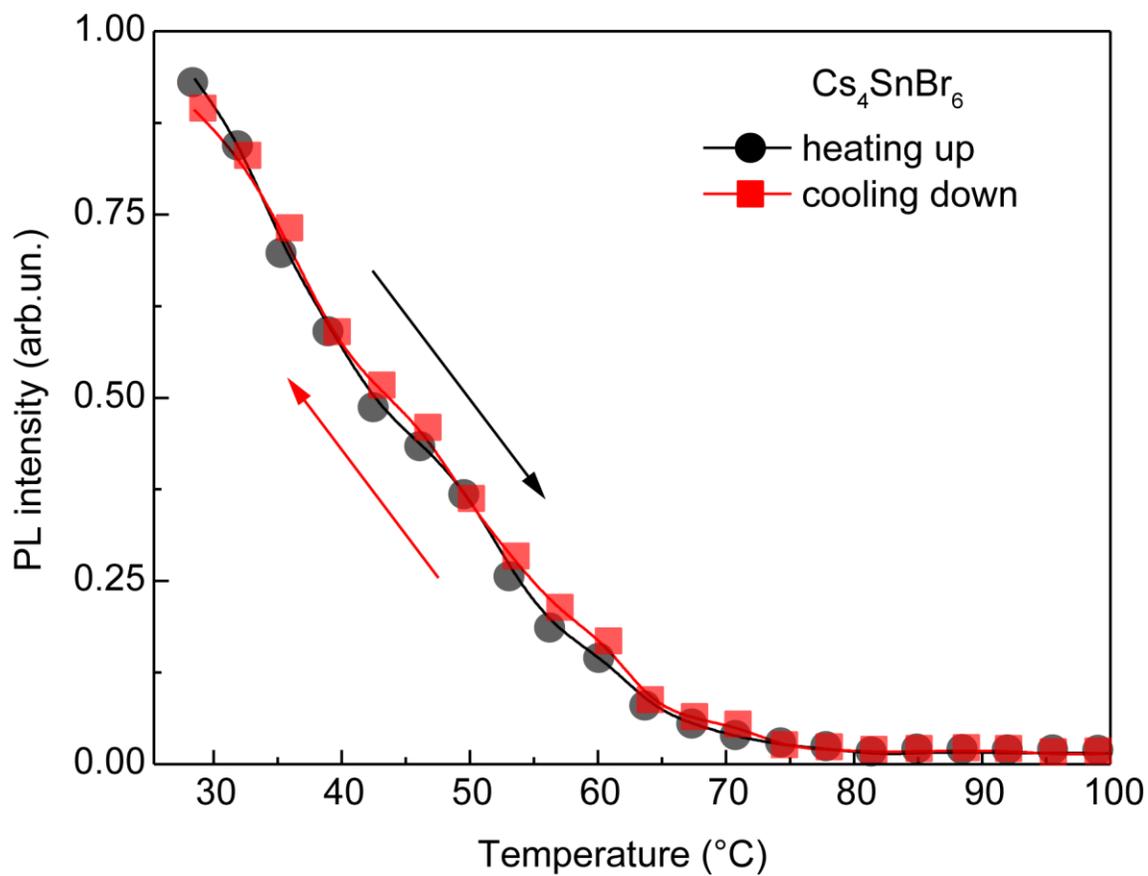

**Supplementary Figure 6.** Reversible thermal quenching for Cs$_4$SnBr$_6$ PL emission.

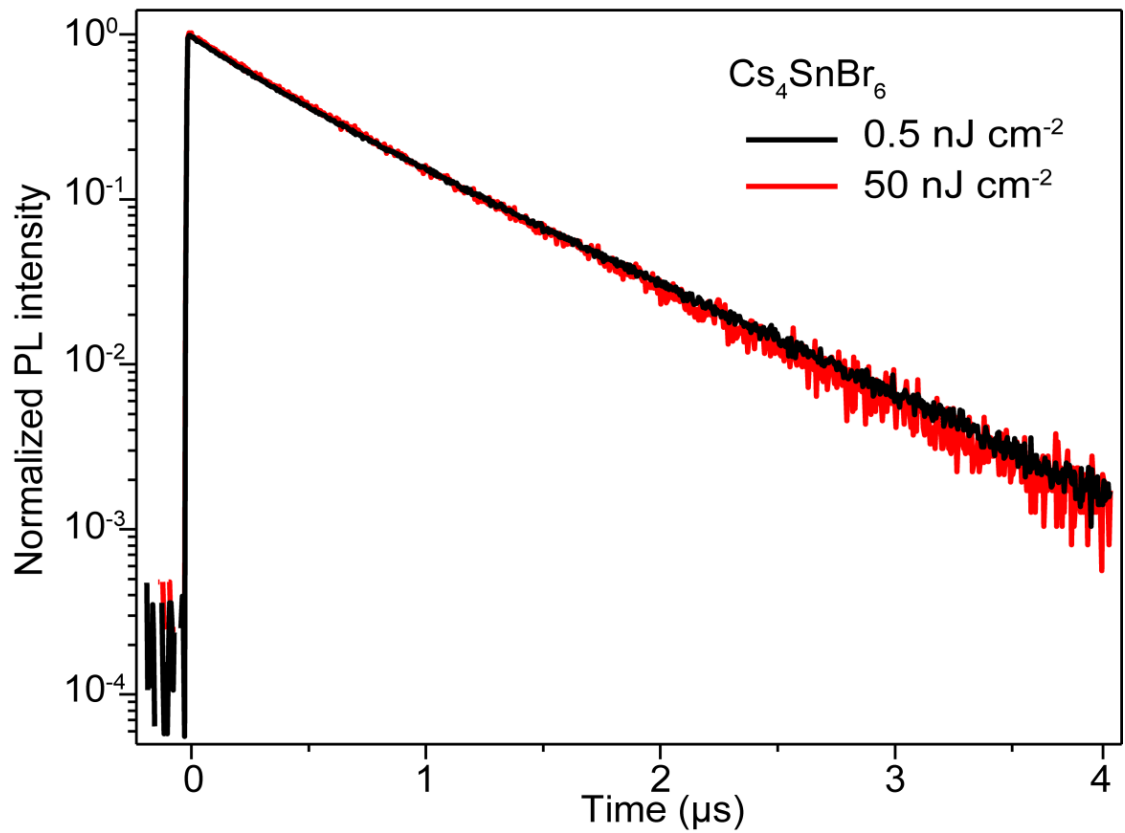

**Supplementary Figure 7.** Time-resolved PL traces for Cs$_4$SnBr$_6$ at different excitation energy densities.

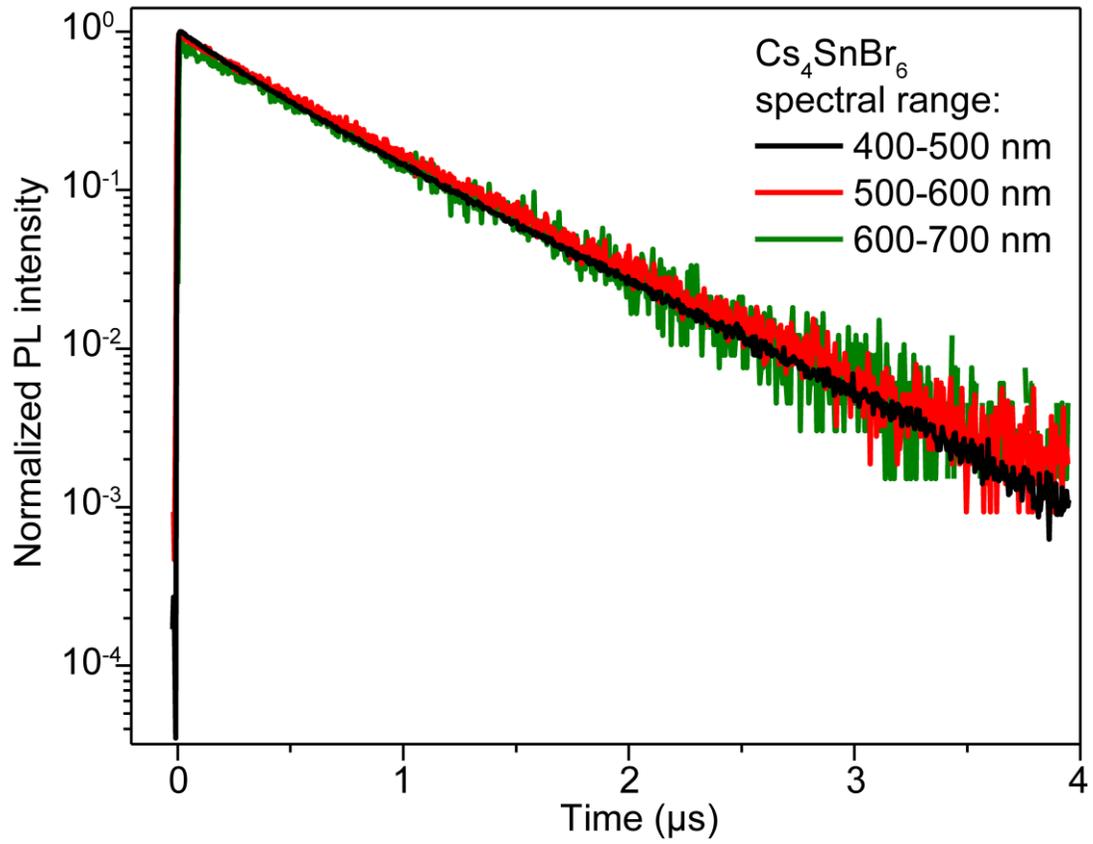

**Supplementary Figure 8.** Time-resolved PL traces for Cs₄SnBr₆ at various emission ranges.

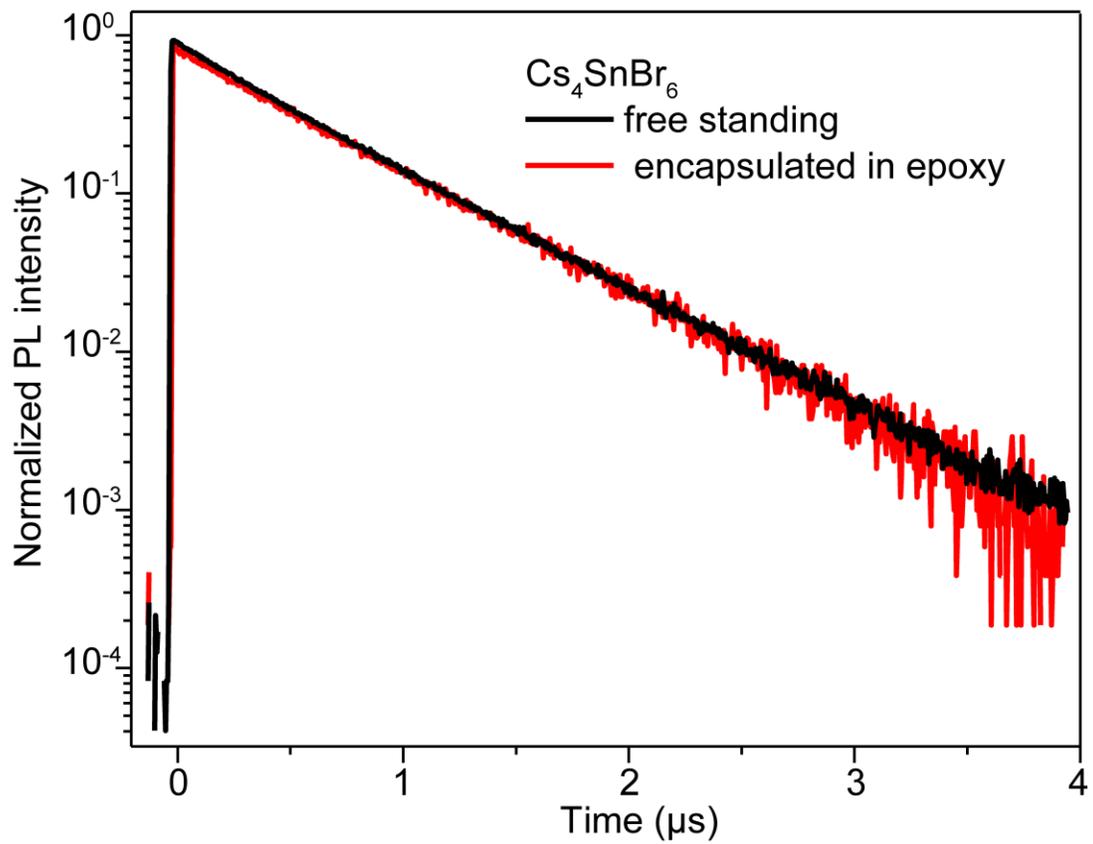

**Supplementary Figure 9.** Time-resolved PL traces for $Cs_4SnBr_6$ under different environmental conditions.

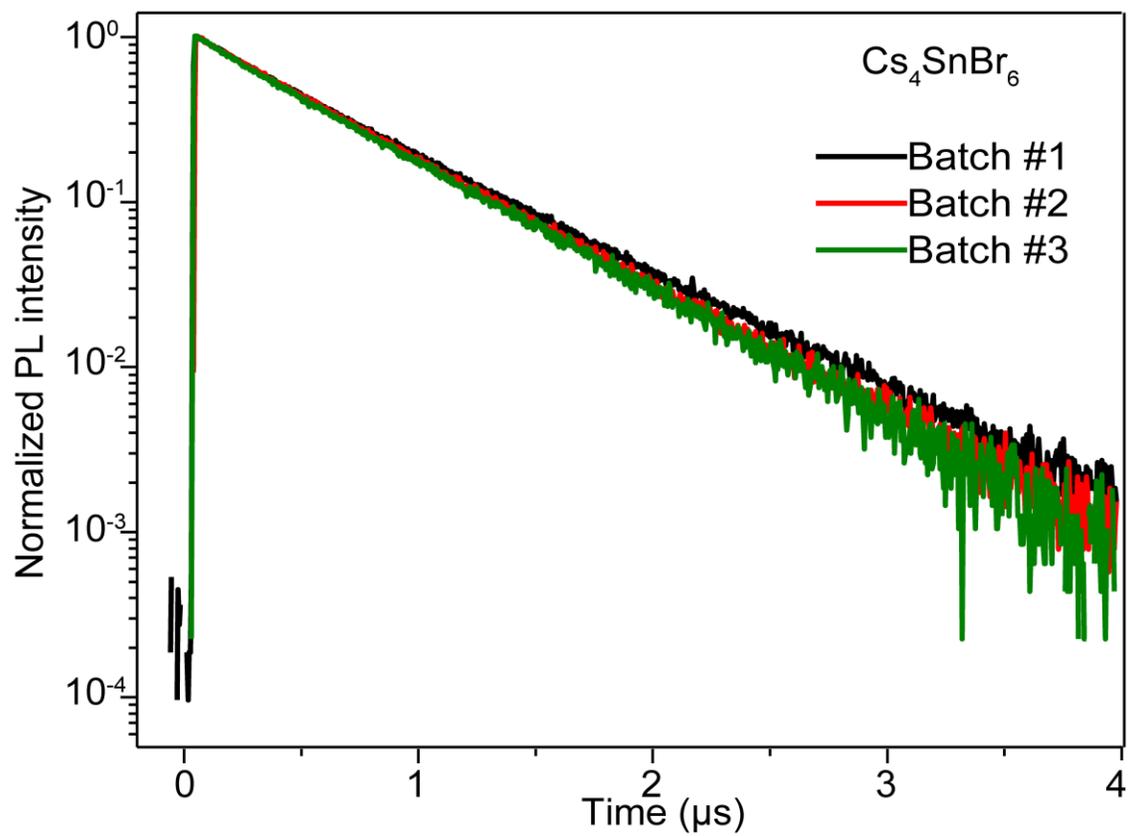

**Supplementary Figure 10.** Time-resolved PL traces of Cs₄SnBr₆ for several synthetic batches.

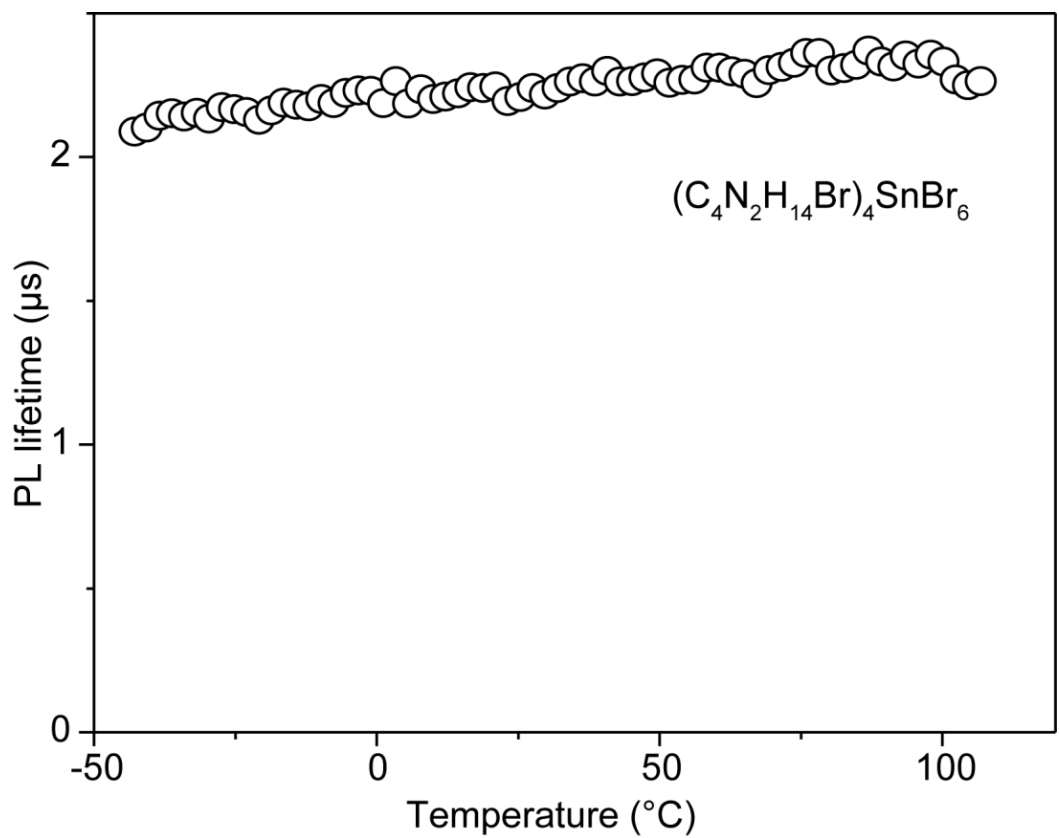

**Supplementary Figure 11.** PL lifetime temperature dependence for $(C_4N_2H_{14}Br)_4SnBr_6$.

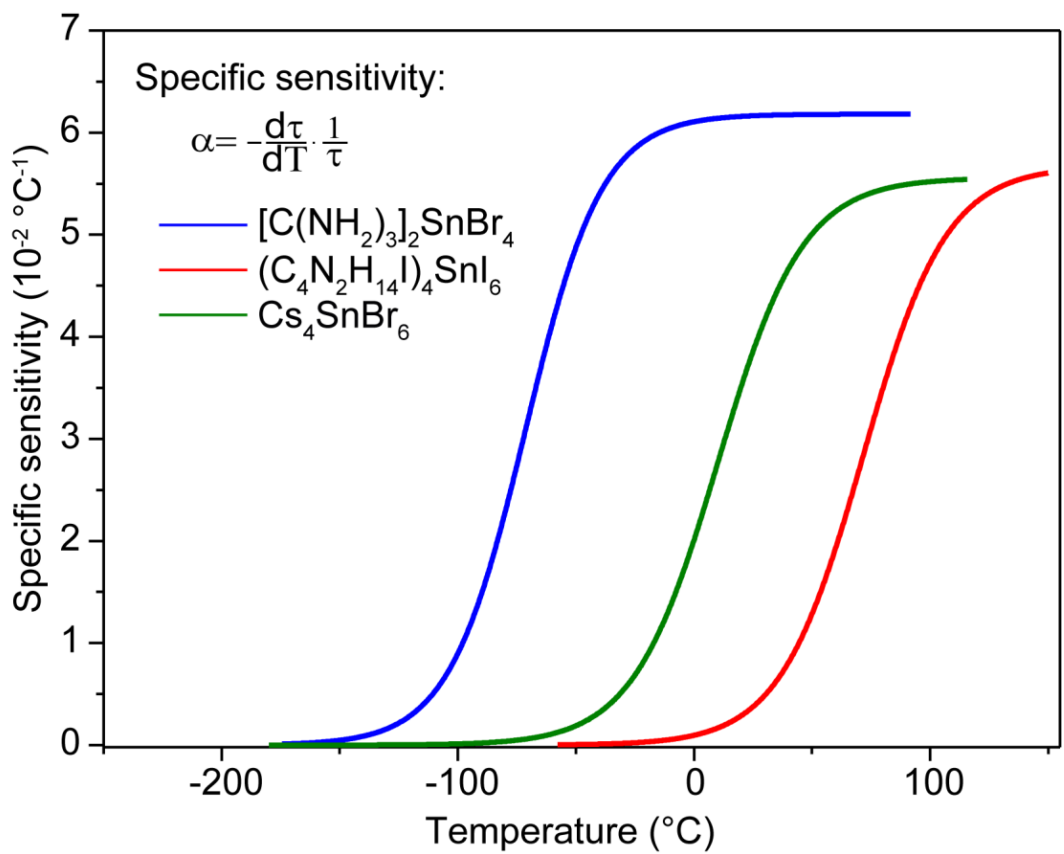

**Supplementary Figure 12.** Specific sensitivity. The dependence of specific sensitivity on temperature for $[C(NH_2)_3]_2SnBr_4$ (blue), $Cs_4SnBr_6$ (green), $(C_4N_2H_{14}I)_4SnI_6$ (red).

**Supplementary Note 1.** Models for the fitting of PL lifetime vs. temperature dependence.

In order to understand the nature of the thermal quenching of emission and the lifetime acceleration in tin-halide luminophores, we fit temperature dependent PL-lifetime data with several suitable models (Figs.S13-S15; Table S1-3). With the first model, we suggested a thermal activation process according to the Mott model, $(T) = \frac{\tau_0}{1+A\cdot e^{-\frac{T_a}{T}}}$, where $\tau_0$ is the intrinsic radiative lifetime, $A$ is a pre-exponential factor (in which $\frac{A}{\tau_0}$ is the non-radiative rate), $T_a$ is the activation temperature, and $T$ is the temperature in K. Although the model agrees with the experimental data, we found that it is difficult to provide a physical interpretation for the fitting parameters: $A$ and $T_a$, which are exceedingly large ($10^7$ and 6000 K, respectively; Fig. S13, Table S1).

Next, we chose the Boltzmann-sigmoid model, which also provided a proper fit for the experimental data according to the equation: $\tau(T) = \frac{\tau_0}{1+e^{\frac{T-T_B}{\Delta T}}}$ where $\tau_0$ is the intrinsic radiative lifetime, $T_B$ and $\Delta T$ are respectively the center and half-width of the temperature sensitive range, and $T$ is the temperature in K (Fig.S14). Despite the fact that the fitting parameters in this model (Table S2) do indeed yield realistic values, the model itself cannot be attributed to any physical quenching process.

Therefore, as a compromise between simplicity and the ability to realistically interpret the model, we chose the exciton-phonon scattering model: $\frac{1}{\tau(T)} = \frac{1}{\tau_0} + \frac{\Gamma_{ph}}{\left(e^{\frac{E_{ph}}{kT}}-1\right)^m}$, where $\tau_0$ is the intrinsic radiative lifetime, $\Gamma_{ph}$ is the exciton-phonon scattering probability, $E_{ph}$ is the phonon energy, $m$ is the number of phonons, $k$ is the Boltzmann constant, and $T$ is the temperature in K (Fig. S15, Table S3).[1]

This model succeeds in providing phonon energies that can be converted to physically relevant activation temperatures (Table S3, column where $E_{ph}$ is shown in units of K). Furthermore, these temperatures agree extremely well with the observed onset of lifetime acceleration in each case. This model, however, predicts a rather high number of phonons involved in the scattering process ($m \sim 10$). Such high values of $m$ might be explained through collective processes that involve high numbers of phonons, but this is also unrealistic given the low probability of such an event occurring. It is possible that there exists an essentially nonlinear temperature dependence in the phonon-exciton interaction, and the self-trapped excitons within such systems require another model to provide a satisfactory description of their thermal behavior.

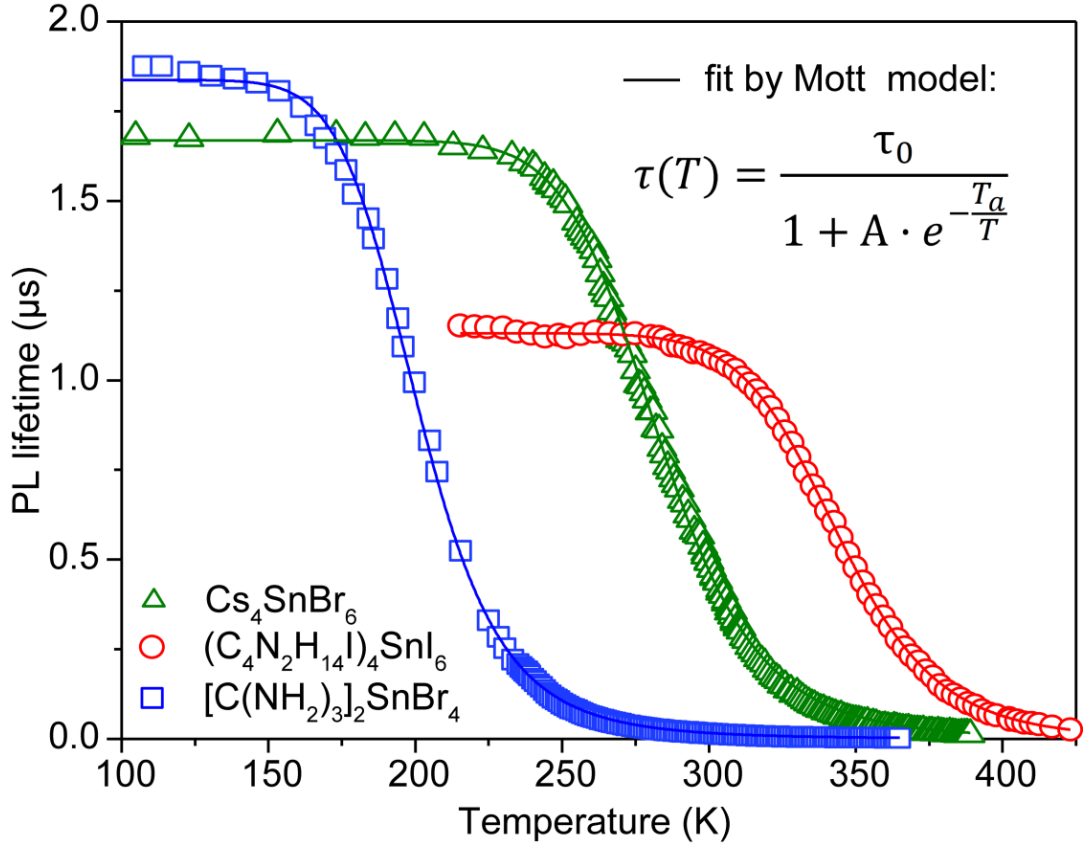

**Supplementary Figure 13.** Fitting by the Mott model. PL lifetime temperature dependence for [C(NH$_2$)$_3$]$_2$SnBr$_4$ (blue squares), Cs$_4$SnBr$_6$ (green triangles), (C$_4$N$_2$H$_{14}$I)$_4$SnI$_6$ (red circles) by fitting with an Mott model (colored lines): $\tau(T) = \frac{\tau_0}{1 + A \cdot e^{-\frac{T_a}{T}}}$.

**Supplementary Table 1**. Fitting parameters for the Mott model.

| Composition | $\tau_0$ | $A$ | $T_a$ | Temperature sensitivity range |
|---|---|---|---|---|
| | ns | | K | °C (K) |
| Cs$_4$SnBr$_6$ | 1669 | $1.9 \cdot 10^7$ | 4759 | −30 – 40 (243 - 323) |
| [C(NH$_2$)$_3$]$_2$SnBr$_4$ | 1837 | $1.5 \cdot 10^6$ | 2863 | −100 – 30 (173 - 243) |
| (C$_4$N$_2$H$_{14}$I)$_4$SnI$_6$ | 1131 | $5.7 \cdot 10^8$ | 6945 | 40 – 110 (313 - 383) |

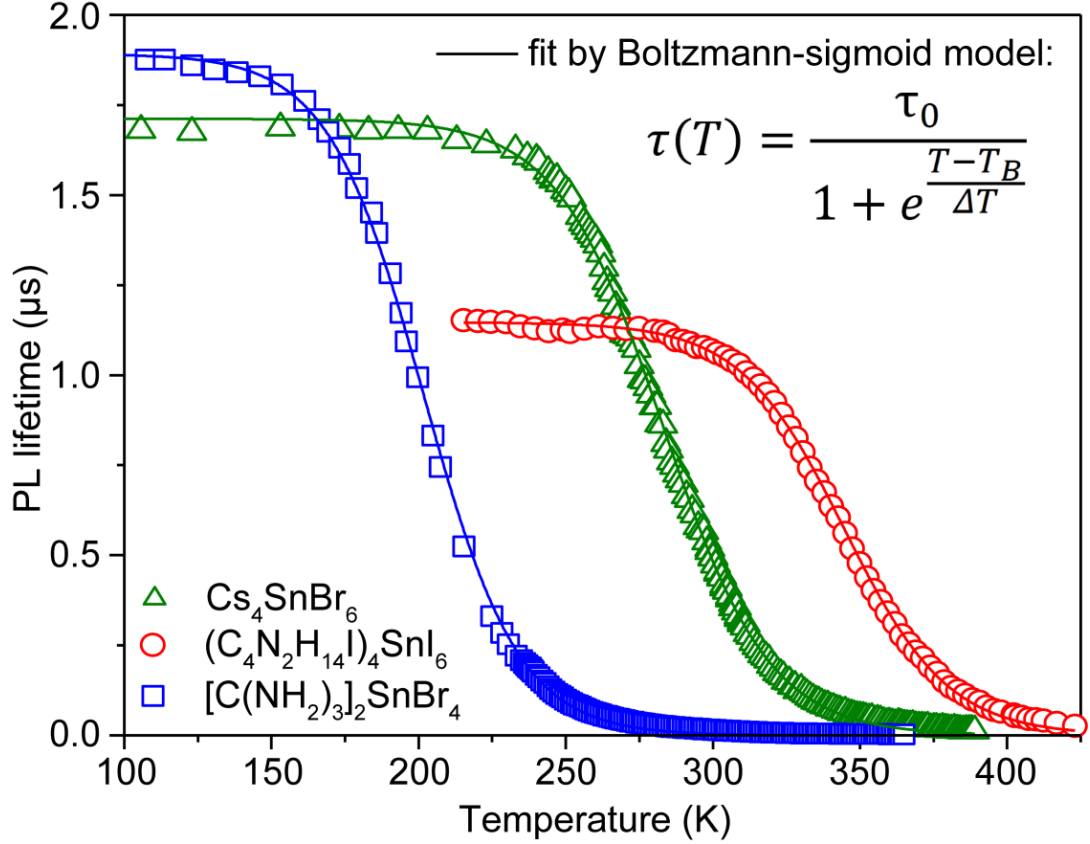

**Supplementary Figure 14.** Fitting by the Boltzmann-sigmoid model. PL lifetime temperature dependence for [C(NH$_2$)$_3$]$_2$SnBr$_4$ (blue squares), Cs$_4$SnBr$_6$ (green triangles), (C$_4$N$_2$H$_{14}$I)$_4$SnI$_6$ (red circles) by fitting with the Boltzmann-sigmoid model (colored lines): $\tau(T) = \frac{\tau_0}{1+e^{\frac{T-T_B}{\Delta T}}}$.

**Supplementary Table 2.** Fitting parameters for Boltzmann-sigmoid model. Temperature sensitivity ranges are determined as T$_B$ ± 2ΔT.

| Composition | $\tau_0$ | $T_B$ | $\Delta T$ | Temperature sensitivity range | $E_{ph}$ |
|---|---|---|---|---|---|
| | ns | K | K | °C (K) | meV |
| Cs$_4$SnBr$_6$ | 1711 | 281 | 18 | -30 – 40 (243 - 323) | 24 |
| [C(NH$_2$)$_3$]$_2$SnBr$_4$ | 1892 | 205 | 17 | -100 – -30 (173 - 243) | 18 |
| (C$_4$N$_2$H$_{14}$I)$_4$SnI$_6$ | 1146 | 345 | 18 | 40 – 110 (313 - 383) | 30 |

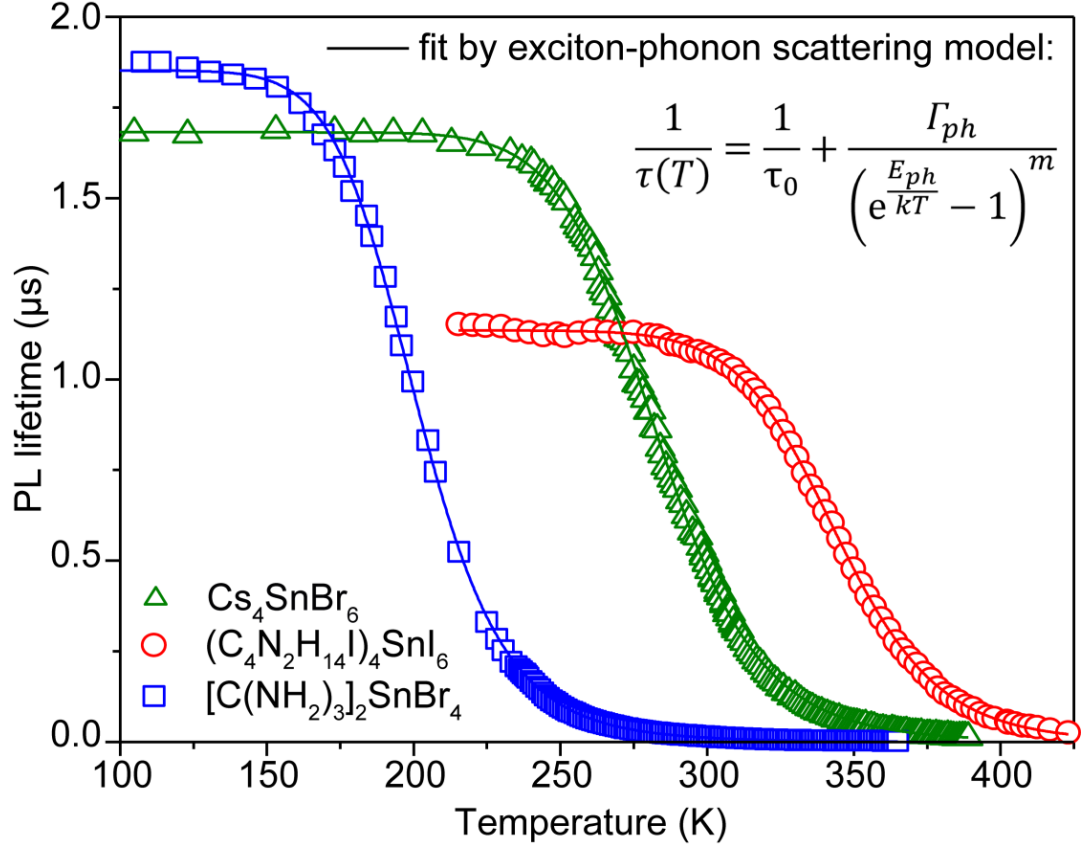

**Supplementary Figure 15.** Fitting by the exciton-phonon scattering model. PL lifetime temperature dependence for [C(NH$_2$)$_3$]$_2$SnBr$_4$ (blue squares), Cs$_4$SnBr$_6$ (green triangles), (C$_4$N$_2$H$_{14}$I)$_4$SnI$_6$ (red circles) by fitting with the exciton-phonon scattering model (colored lines):

$\frac{1}{\tau(T)} = \frac{1}{\tau_0} + \frac{\Gamma_{ph}}{\left(e^{\frac{E_{ph}}{kT}} - 1\right)^m}$.

**Supplementary Table 3**. Fitting parameters for the exciton-phonon scattering model.

| Composition | $\tau_0$ | $\Gamma_{ph}$ | $E_{ph}$ | | $m$ | Temperature sensitivity range |
|---|---|---|---|---|---|---|
| | ns | $10^{-6}$ s$^{-1}$ | meV | K | | °C (K) |
| Cs$_4$SnBr$_6$ | 1681 | 0.053 | 21.3 | 247 | 10.7 | -30 – 40 (243 - 323) |
| [C(NH$_2$)$_3$]$_2$SnBr$_4$ | 1851 | 0.03 | 15.6 | 181 | 8.7 | -100 – -30 (173 - 243) |
| (C$_4$N$_2$H$_{14}$I)$_4$SnI$_6$ | 1135 | 0.13 | 25.2 | 292 | 13 | 40 – 110 (313 - 383) |

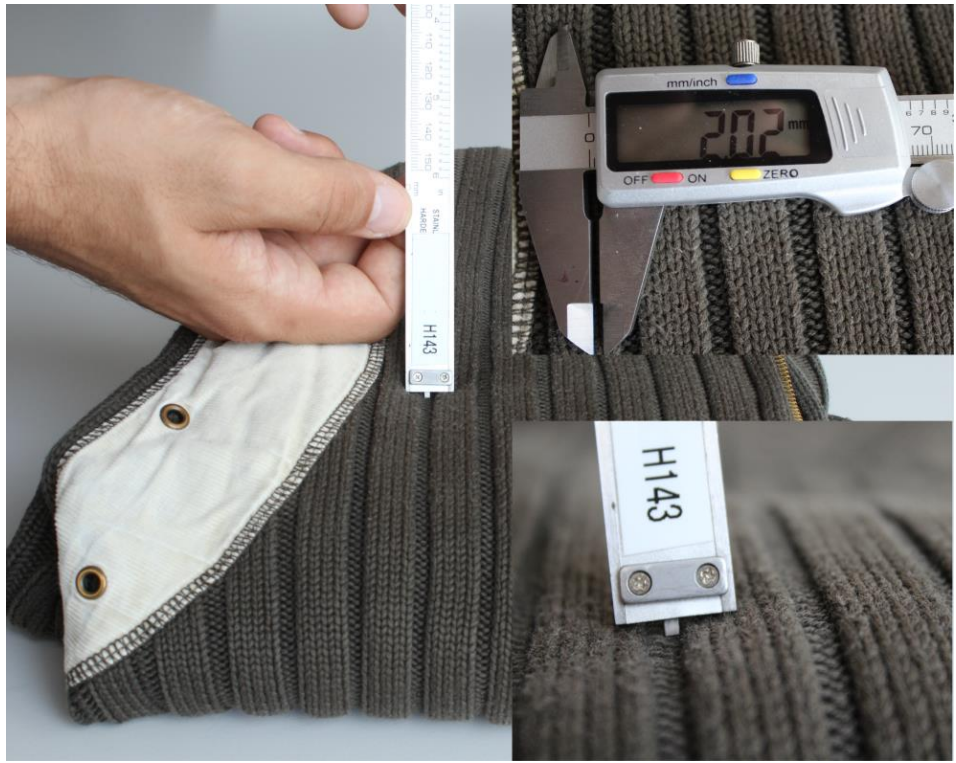

**Supplementary Figure 16**. Photographs of the depth standard for ToF imaging in Fig.4a.

**Supplementary Note 2.** Basic principles of ToF-FLI.

Depth images such as those produced by the ToF sensor in the Kinect 2.0 from Microsoft Corp. are measured by the phase shift $\Delta\phi$ (equivalent to the delay time) of the reflected light relative to the IR reference beam due to light propagation $x$ (Supplementary Fig.S17a). For the depth imaging we take into account the fact that the reflection is an immediate process on the contrary to PL that has an intrinsic delay associated with the PL lifetime. The ToF-measurement of the delay caused by the luminophore's PL lifetime $\tau$ makes the use of ToF–FLI possible (Supplementary Fig.S17b). To do so, we switched to UV excitation (as an analog to the IR reference beam used in the Kinect 2.0) and rejected any scattered or reflected UV light by optical cut-off filters to exclusively record the PL delay. However, an additional reference measurement without optical filters (to register UV light reflected by the sample) is required to determine the initial delay time $\Delta t_0$. This delay time occurs as a result of the optical path of the measurement scheme as well as variations which result from inhomogeneities in a sample's topography. This must be correctly accounted for to precisely determine the PL-lifetime, $\tau$.

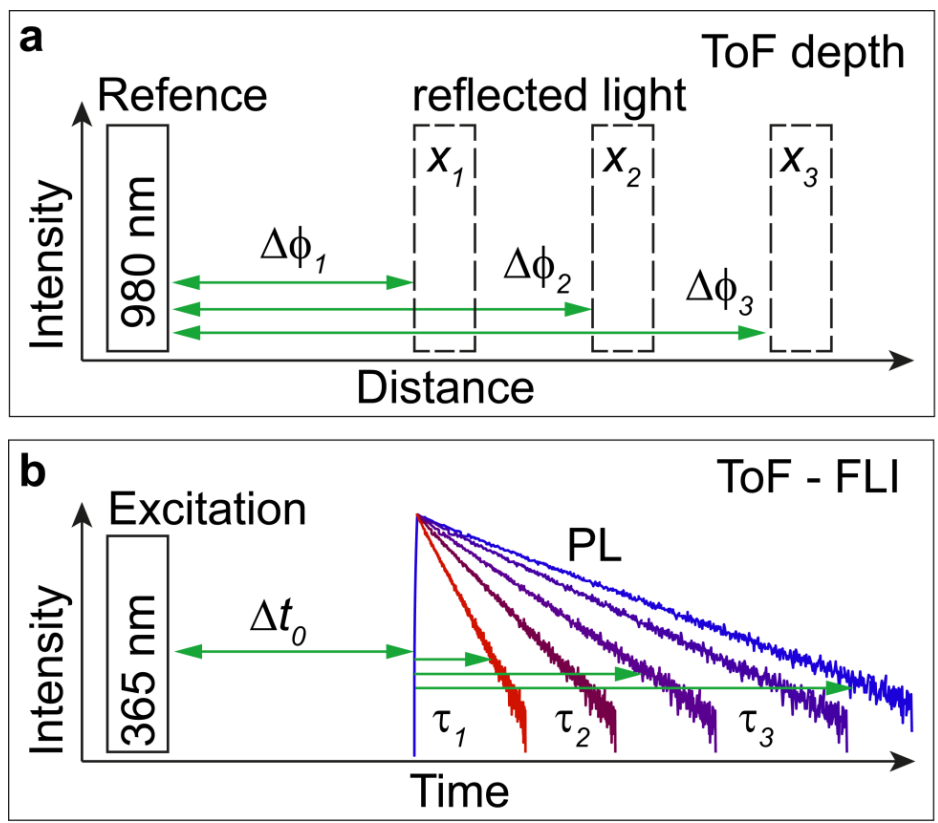

**Supplementary Figure 17.** Scheme describing optical ToF measurements. (a) The estimation of distance *x* by the measured phase shift *Δϕ*. (b) Similar ToF hardware principles apply to the estimation of PL decay.

**Supplementary Note 3.** Frequency domain PL lifetime measurement by phase-shift.

ToF-FLI is based on a mathematical model where the observed PL emission trace is shown as a convolution of a harmonically modulated excitation signal with a mono-exponential emission relaxation decay:

$$I_{PL}(t) = \int_0^t B \cdot \left(1 + cos\left(2\pi\nu x + \frac{\pi}{2}\right)\right) \cdot e^{-\frac{t-x}{\tau}} dx, \tag{1}$$

where $B$ is a scaling coefficient, $\nu$ is the excitation modulation frequency, and $\tau$ is an exponential decay parameter. The result of convolution (1) is a harmonically oscillating trace that has a certain phase delay $\phi$ from the excitation trace (Fig.4b). In the case of a mono-exponential decay the result of the phase delay is determined[2] by:

$$\tau = \frac{\tan(\Delta\phi)}{2\pi\nu} \tag{2}$$

The working principle for a ToF image sensor is based on the acquisition of four, phase-locked images at 0°, 90°, 180° and 270° phase delay with respect to the excitation signal ($I_0$, $I_1$, $I_2$ and $I_3$ on Fig.4b).[3] From these four images, the spatial distribution of the PL intensity $I$, modulation index $M$, and phase angle $\Delta\phi$ can be computed:

$$\Delta\phi = \arctan\left(\frac{I_0 - I_2}{I_3 - I_1}\right) \tag{3}$$

$$M = 2\frac{\sqrt{(I_1 - I_3)^2 - (I_0 - I_2)^2}}{I_0 + I_1 + I_2 + I_3} \tag{4}$$

$$I = \frac{I_0 + I_1 + I_2 + I_3}{4} \tag{5}$$

The lifetime is then calculated using Eq. (2). In the ToF-FLI prototype used in this study (Fig. S18), each pixel of the imaging chip is capable of performing this measurement in parallel, thus circumventing the need for bulky and expensive scanning systems.

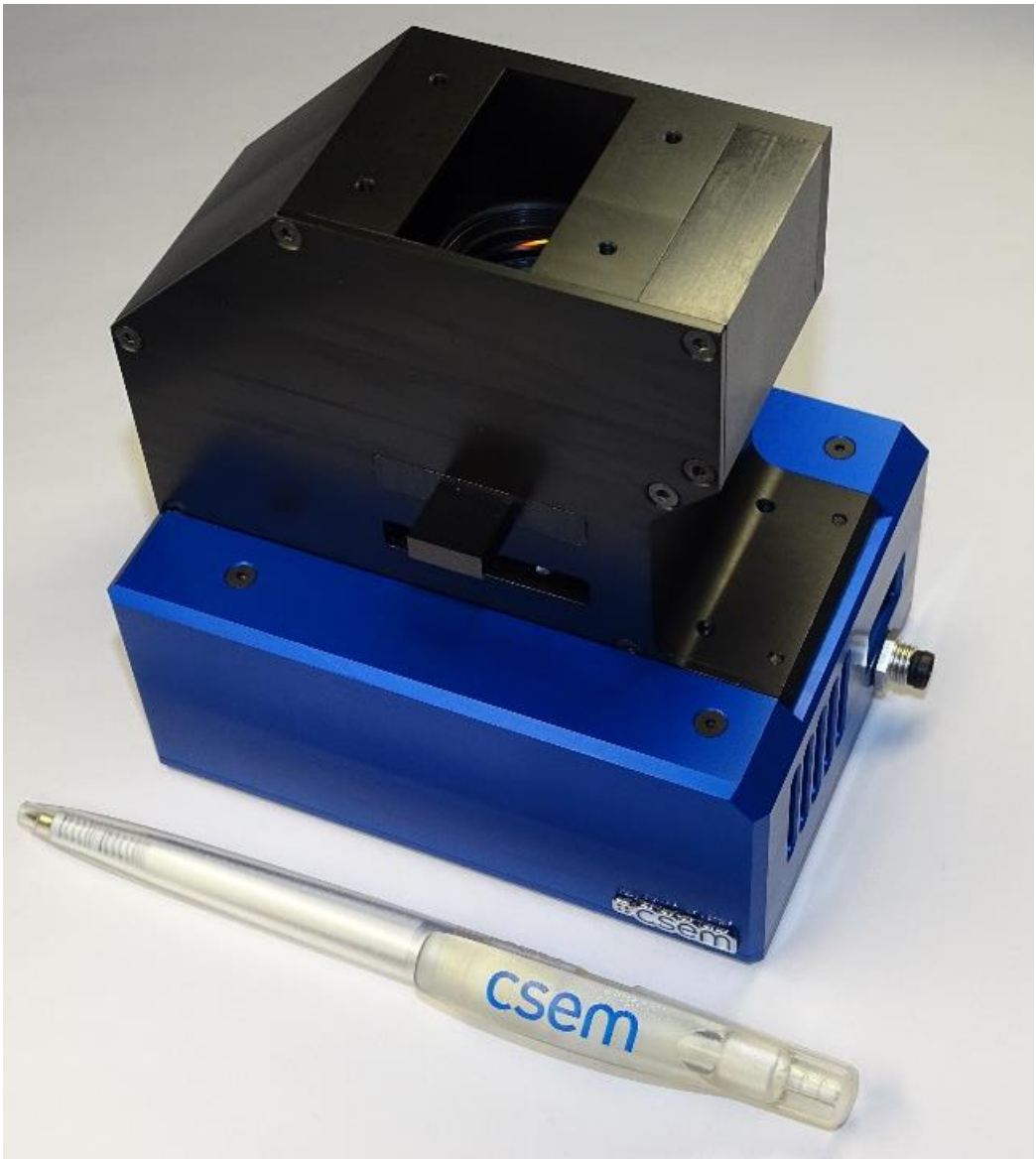

**Supplementary Figure 18.** Compact stand-alone ToF-FLI prototype developed by CSEM (Switzerland) for real-time, wide-field fluorescence lifetime imaging in the nano- to micro-second range.

**Supplementary Note 4**. Key specifications of ToF-FLI image sensor used in the setup:

- Number of pixels: 256 x 256 pixels

- Maximum frame rate at full resolution: 100 fps

- Analog outputs (no on-chip ADC)

- Pixel size: 6.3 x 6.3 μm → Active area size 1.6 mm x 1.6 mm

- Maximum demodulation frequency: 4 x 20 MHz = 80 MHz

- Quantum efficiency (QE): 0.43@600nm, 0.21@800nm, 0.14@850nm

- Fill factor: 14 %

- Dark current: 40 e-

- Pixel saturation capacitance: 9 ke-

- Maximum SNR: 39.9 dB

- Dynamic range: 49.9 dB

Detailed description of image sensor architecture can be found in Supplementary Ref. [4].

**Supplementary Figure 19.** A thermographic image of a patterned ITO glass slide with a bolometric camera. The sample was heated with a passing electrical current, and acquired with a commercial Seek Thermal Compact Pro™ LWIR bolometry camera equipped with a ZnSe lensed macro-objective.

**Supplementary Video 1.** Thermographic video.

The dynamics of temperature change in the sample and heat transfer through the substrates due to brief contact with a hot soldering pin (the temperature of pin apex was approx. 120 °C). The sample is $(C_4N_2H_{14}I)_4SnI_6$ powder encapsulated between two relatively thick (1mm) glass substrates.

**Supplementary References**